\documentclass[]{aastex631}
\usepackage{subfigure}
\usepackage{graphicx,multirow} 
\usepackage{booktabs}
\usepackage{float}
\usepackage{ulem}
\usepackage{threeparttable}

\newcommand{\ssst}{\scriptscriptstyle}
\newcommand{\E}[1]{\times 10^{#1}}

\newcommand{\lt}{\left}       \newcommand{\rt}{\right}

\newcommand{\s}{\,{\rm s}}      \newcommand{\ps}{\,{\rm s}^{-1}}
    
\newcommand{\cm}{\,{\rm cm}}    \newcommand{\km}{\,{\rm km}}

\newcommand{\erg}{\,{\rm erg}}        \newcommand{\K}{\,{\rm K}}
    \newcommand{\keV}{\,{\rm keV}}

\newcommand{\nel}{n_{e}}        
\newcommand{\ti}{t_{\rm i}}
   
\newcommand{\Rs}{R_{\rm s}}         \newcommand{\Vs}{V_{\rm s}}
\newcommand{\nH}{n_{\ssst\rm H}}        \newcommand{\mH}{m_{\ssst\rm H}}
\newcommand{\OVIIr}{O\,VII\,He$\alpha$\,r}
\newcommand{\OVIIIa}{O\,VIII\,Ly$\alpha$}
\newcommand{\OVIIIb}{O\,VIII\,Ly$\beta$}
\newcommand{\taum}{\tau_{\nu_0m}}

\newcommand{\inlet}[1]{{\color{red}{#1}}}

\begin{document}

\title{A Monte-Carlo Simulation on Resonant Scattering of X-ray Line Emission in Supernova Remnants
 }

\correspondingauthor{Yang Chen}
\email{ygchen@nju.edu.cn}

\author[0000-0001-7571-2318]{Yiping Li}
\affiliation{Department of Astronomy, Nanjing University, 163 Xianlin Avenue, Nanjing 210023, People's Republic of China}

\author[0000-0002-0747-0078]{Gao-Yuan Zhang}
\affiliation{Departamento de Astronomía, Universidad de Concepción, Chile}

\author[0000-0002-4753-2798]{Yang Chen}
\affiliation{Department of Astronomy, Nanjing University, 163 Xianlin Avenue, Nanjing 210023, People's Republic of China}
\affiliation{Key Laboratory of Modern Astronomy and Astrophysics, Nanjing University, Ministry of Education, People's Republic of China}

\author[0000-0001-9671-905X]{Lei Sun}
\affiliation{Department of Astronomy, Nanjing University, 163 Xianlin Avenue, Nanjing 210023, People's Republic of China}
\affiliation{Key Laboratory of Modern Astronomy and Astrophysics, Nanjing University, Ministry of Education, People's Republic of China}

\author[0000-0002-8329-6606]{Shuinai Zhang}
\affiliation{Purple Mountain Observatory, Chinese Academy of Sciences, People's Republic of China}
\affiliation{Key Laboratory of Dark Matter and Space Astronomy, CAS, People's Republic of China}

\begin{abstract}

Resonant scattering (RS) of X-ray line emission in supernova remnants (SNRs) may modify the observed line profiles and fluxes and has potential impact on estimating the physical properties of the hot gas and hence on understanding the SNR physics,
but has not been theoretically modeled ever.
%
Here we present our Monte-Carlo simulation of RS effect on X-ray resonant-line emission, typified by \OVIIr\ line, from SNRs.
We employ the physical conditions characterized by the Sedov-Taylor solution and some basic parameters similar to those in Cygnus Loop.
We show that the impact of RS effect is most
significant near the edge of the remnant.
The line profiles are predicted to be asymmetric because of different temperatures and photon production efficiencies of the expanding gas at different radii.
We also predict the surface brightness of the line emission would decrease in the outer projected region but is slightly enhanced in the inner.
The G-ratio of the O\,VII He$\alpha$ triplet can be effectively elevated by RS in the outer region.
We show that RS effect of the \OVIIr\ line in the southwestern boundary region of Cygnus Loop is non-negligible. The observed O\,VII G-ratio $\sim$1.8 of the region could be achieved with RS taken into account for properly elevated O abundance from the previous estimates.
Additional simulation performed for the SNRs in ejecta-dominated phase like Cas A shows that RS in the shocked
ejecta may have some apparently effects on the observational properties of oxygen resonant lines.


\end{abstract}

\keywords{Radiative transfer (1335); Supernova remnants (1667); X-ray astronomy (1810); Atomic spectroscopy (2099)}


\section{Introduction}\label{sec: intro}
High-resolution X-ray spectroscopy provides detailed plasma diagnostics of supernova remnants (SNRs) and facilitate insights into the radiation processes and shock physics of the interior hot plasma.
In the spectral analyses, resonant scattering (RS) of line emission often plays a non-negligible role.

H-like and He-like ions give the most prominent line features in the thermal X-ray spectra, and the He-like ion line ratios prove to be valuable tools in high-resolution spectral analysis \citep[e.g.,][]{2010SSRv..157..103P}.
However, there are unresolved issues 
in modeling 
these lines, such as high G-ratios, the ratio of the intercombination (i) line and forbidden (f) line to the resonance (r) line \citep{1969MNRAS.145..241G}, of He-like triplet in some SNRs \citep[e.g.,][]{2019ApJ...871..234U,2020ApJ...900...39S,2022PASJ...74..757K,2022ApJ...933..101T}. 
So far two candidate mechanisms have been suggested for interpretation.
One is charge exchange (CX)
between ions and neutral atoms \citep[e.g.,][]{2022ApJ...933..101T,2012ApJ...756...49K}, 
which usually occurs when the hot plasma interacts with cold gas 
and strengthens the f line \citep[e.g.,][]{2019ApJ...885..157Z}. 
Another one is RS \citep[e.g.,][]{2008PASJ...60..521M,2020ApJ...897...12A}, which changes the r line flux when the plasma is optically thick to the line emission \citep[e.g.,][]{2010SSRv..157..103P}.
In this work, we will focus on the latter mechanism. 

RS is the process of a resonant-line photon encountering an atom or ion, in which the particle absorbs the photon and is soon deexcitated to the original energy level by re-emitting another photon in a random direction.
The process looks like a photon scattering event, but the energy of the scattered photon is shifted usually due to the thermal and turbulent random motions of the particle. Such resonant ``scattering'' events usually happen in the lines with large oscillation strength 
(such as Ly$\alpha$, He$\alpha$ r line, etc.) 
and may, especially for optically thick cases, alter the width and profile of the line and distort the surface brightness distribution in an extended source \citep[see, e.g.,][]{1987SvAL...13....3G, 2018ApJ...861..138C}.
In X-ray spectroscopic diagnostics, neglect of RS effect could lead to a bias of estimating physical properties of the X-ray emitting gas, such as temperature, metal abundance, and ionization parameter, thus misleading the inference of the SNR physics and the SNR progenitors.

RS has been noted to play potential roles in the X-ray emission in a few SNRs, such as Cas\,A \citep{1995A&A...302L..13K}, Cygnus Loop \citep{2008PASJ...60..521M}, N49 \citep{2020ApJ...897...12A}, and N132D \citep{2020ApJ...900...39S}. 
For example, Suzaku X-ray observation found significant RS optical depth for the bright emission of O\,VII resonant lines in the northeastern rim of the SNR, which was suggested implying a factor of $>10\%$ underestimation of the abundance of O \citep{2008PASJ...60..521M}. An XMM-Newton RGS X-ray observation discovered a high forbidden-to-resonant line ratio of the O\,VII line in the southwestern edge of the SNR and suggested that RS effect can not be ruled out \citep{2019ApJ...871..234U}.
%
%
However, there have been few modeling studies of the impact of the RS effect on the X-ray observations of SNRs in literature as yet, 
while this effect has often been investigated in diffuse hot plasma of elliptical galaxies and clusters of galaxies \citep[see][for detailed references]{2018ApJ...861..138C}. 
We make an attempt in modeling the RS effect on the X-ray resonant line emission of SNRs using Monte-Carlo (MC) simulation.
The calculation is performed by adapting the scheme formerly developed for the RS in galactic bulges \citep{2018ApJ...861..138C}.
Our modeling for the RS effect is applied to the case of high O\,VII G-ratio typified by the Cygnus Loop SNR, which has been observed in \OVIIr\ line in some detail.



\section{RS modeling} \label{sec: RS modeling}
Our consideration is focused on in the adiabatic phase of the SNR, with a simple extension to the case of ejecta-dominated case.
We assume a spherically symmetric geometry in a homogeneous and uniform interstellar medium (ISM). 
Although the ionization parameter $n_{\rm e}\ti$ can be different at the different locations, we assume a single $n_{\rm e}\ti$ across the whole remnant.
We assume the electron temperature is equal to the ion temperature 
\citep{1993A&AS...97..873K} and the both are denoted as $T$ in this paper.

\subsection{Plasma Distribution Interior to the SNR}\label{sec: ST Distribution}

\begin{figure}[ht!]
\centerline{
\includegraphics[scale=0.6]{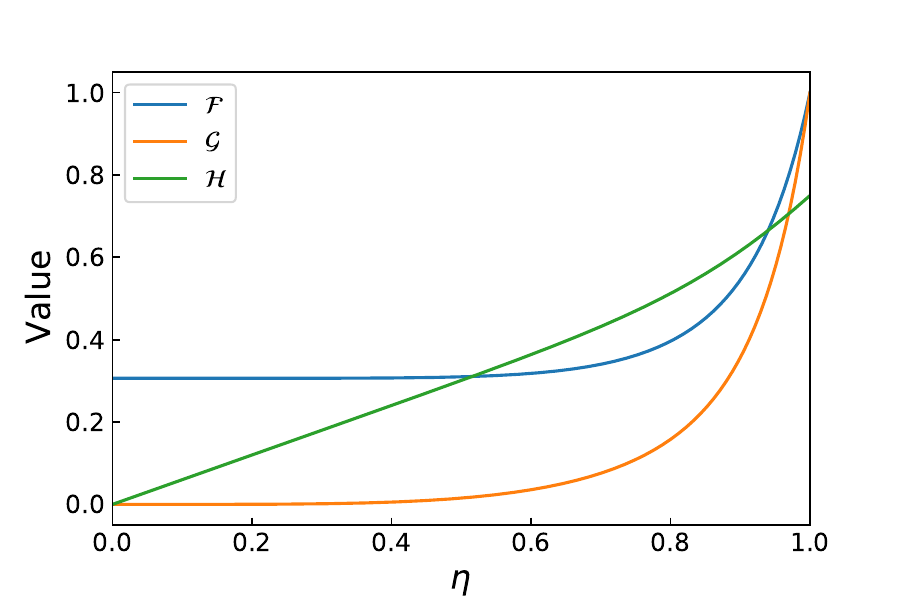}
}
\caption{The distribution of gas pressure, density, and velocity with radius interior to the remnant, as given by the Sedov-Taylor solution.} 
\label{fig: STsolve}
\end{figure}

We adopt Sedov-Taylor's self-similar solution \citep{1959sdmm.book.....S} to described the dynamic evolution and interior plasma distribution of the SNR in the adiabatic phase (also named Sedov-Taylor (ST) phase). 
The case of the ejecta-dominated phase will be discussed in \S\ref{sec:Simple discussion on RS in ejecta-dominated SNR} and Appendix \ref{app: The RS of Ejecta}. 
In Sedov-Talor phase phase, the gas gets cold so that the emission of O\,VII triplet will be observed obviously related to the gas temperature shown in Figure\,11 in \citet{2012A&ARv..20...49V}. The bulk velocities will also be smaller so that the RS effect will become significant in the phase \citep{1995A&A...302L..13K}.
The radius of the remnant, $R_{\rm s}$, evolves with time $t$ as
\begin{equation}
 R_{\rm s}=\left(\frac{\xi E}{\rho_0}\right)^{1/5}t^{2/5},   
\end{equation}
where $E$ is the supernova explosion energy, $\rho_0$ the masss density of the preshock gas, and $\xi=2.026$.
The SNR expands at a velocity
$V_{\rm s}=2R_{\rm s}/(5t)$.
With the radius $r$ scaled as $\eta=r/R_{\rm s}$ and the gas pressure $p$, density $\rho$, and bulk motion velocity $v_{\rm bulk}$ nondimensionalized as
$\mathcal{F}(\eta)=p(\eta)/p_{\rm s}$,
$\mathcal{G}(\eta)=\rho(\eta)/\rho_{\rm s}$,
$\mathcal{H}(\eta)=v_{\rm bulk}(\eta)/(3V_{\rm s}/4)$,
where $p_{\rm s}$ and $\rho_{\rm s}$ are the postshock pressure and density, respectively,
the numerical solution of them can be represented by the curves plotted in Figure\,\ref{fig: STsolve}.
The gas temperature and number density are given by
\begin{equation} \label{eq: temperature}
    T(\eta)=\frac{3}{16}\frac{\bar{\mu} m_p V_s^2}{k}\frac{\mathcal{F}(\eta)}{\mathcal{G}(\eta)},
\end{equation}
\begin{equation}\label{eq: nH}
    n_{\rm H}(\eta)=4\mathcal{G}(\eta)n_0,
\end{equation}
respectively, where $k$ is the Boltzmann constant, $\bar{\mu}$ is the average atom weight ($\approx 0.61$ for ionized gas), $m_p$ is the proton mass, and $n_0$ is the number density of the preshock hydrogen atoms. Here an adiabatic index of 5/3 has been assigned.

The ion number density of element $Z$ is related to hydrogen density as: 
\begin{equation}\label{eq:nion}
n_{\rm ion}(\eta,Z)=\zeta(Z)A(Z)f_{\rm ion}(T(\eta))n_{\rm H}(\eta), 
\end{equation}
where $\zeta(Z)$ is the elemental abundance relative to solar (the solar abundances are adopted from \citet{1989GeCoA..53..197A}), $A(Z)$ is the solar photosphere elemental abundance with respect to hydrogen, and $f_{\rm ion}(T(\eta))$ is the ion fraction of the element as a function of the gas temperature $T$ and ionization timescale $\nel\ti$ (but the latter is fixed on the assumption of single $\nel\ti$), calculated here with pyAtomDB \citep{2020Atoms...8...49F}.


\subsection{Radiative Transfer in Plasma}

In the {\it plasma's comoving (or local) reference frame} in which the velocity of bulk motion is zero, with turbulent motion ignored, the velocity dispersion of the ions at dimensionless radius $\eta$ are dominated by thermal motion: $\sigma_{v}(\eta)=\left[kT(\eta)/(\mu_am_p)\right]^{1/2}$ (with $\mu_a$ is the atomic weight).
The distribution profile of the seed photons with frequency $\nu$ is
\begin{equation}\label{eq: line profile}
    P_\nu(\nu,\eta)d\nu=\frac{1}{\sqrt{2\pi}\sigma_\nu(\eta)}\exp\left[-\frac{1}{2}\left(\frac{\nu-\nu_0}{\sigma_\nu(\eta)}\right)^2\right]d\nu.
\end{equation}
where $\nu_0$ is the frequency of the line center and $\sigma_\nu=\nu_0 \sigma_v(\eta)/c$ is the standard deviation in frequency. The RS cross section of the photons is
\begin{equation}\label{eq: cross section}
    s(\eta,x)=\frac{\sqrt{\pi}e^2}{m_ec\,\Delta\nu_D(\eta)}f_{\rm lu}H(a(\eta),x(\eta)),
\end{equation}
where $\Delta\nu_D(\eta)=\nu_0\left[\sqrt{2}\sigma_v(\eta)/c\right]$ is the Doppler width, $x(\eta)\equiv(\nu-\nu_0)/\Delta\nu_D(\eta)$ is the dimensionless frequency shift, $f_{\rm lu}$ is the oscillation strength of the lower to upper level transition, $a(\eta)=\Gamma/[4\pi\Delta\nu_D(\eta)]$ is the Voigt parameter (with $\Gamma$ the damping width, equal to the spontaneous emission rate $A_{\rm ul}$), and $H(a(\eta),x(\eta))$ is the Voigt function. The function can be well approximated to be comprised of the contribution of the line core and line wing, namely \citep[see][]{2018ApJ...861..138C}
\begin{equation}\label{eq: Voigt function}
    H(a(\eta),x(\eta))\approx \exp(-x^2(\eta))+\frac{a(\eta)}{\pi^{1/2}x^2(\eta)}[1-\exp(-x^2(\eta))],
\end{equation}
respectively. In the local reference frame, the frequency of the scattered photon is given by
\begin{equation} \label{eq:ph_en}
x_f\approx x_i-\frac{{\bf v}_a\cdot{\bf k}_i}{\sqrt{2}\sigma_v}
       +\frac{{\bf v}_a\cdot{\bf k}_f}{\sqrt{2}\sigma_v}
       +g\,({\bf k}_i\cdot{\bf k}_f-1),
\label{e:energy}
\end{equation}
where ${\bf k}$ is the unit vector of the photon's propagating direction,
the subscripts ``$i$" and ``$f$" denote the quantities of
the incident and scattered photons, respectively, ${\bf v}_a$ is the velocity of the atom (or ion) in the local reference frame, and $g=h\Delta\nu_D/(2\mu_a m_p\,\sigma_v^2)$
is the average fraction of energy transferred per scattering
to the ion due to its recoil. It is negligible by orders of magnitude for the O\,VII~He$\alpha$ resonant lines of interest in the gas.

\begin{figure*}[ht!]
\centerline{
\includegraphics[scale=0.3]{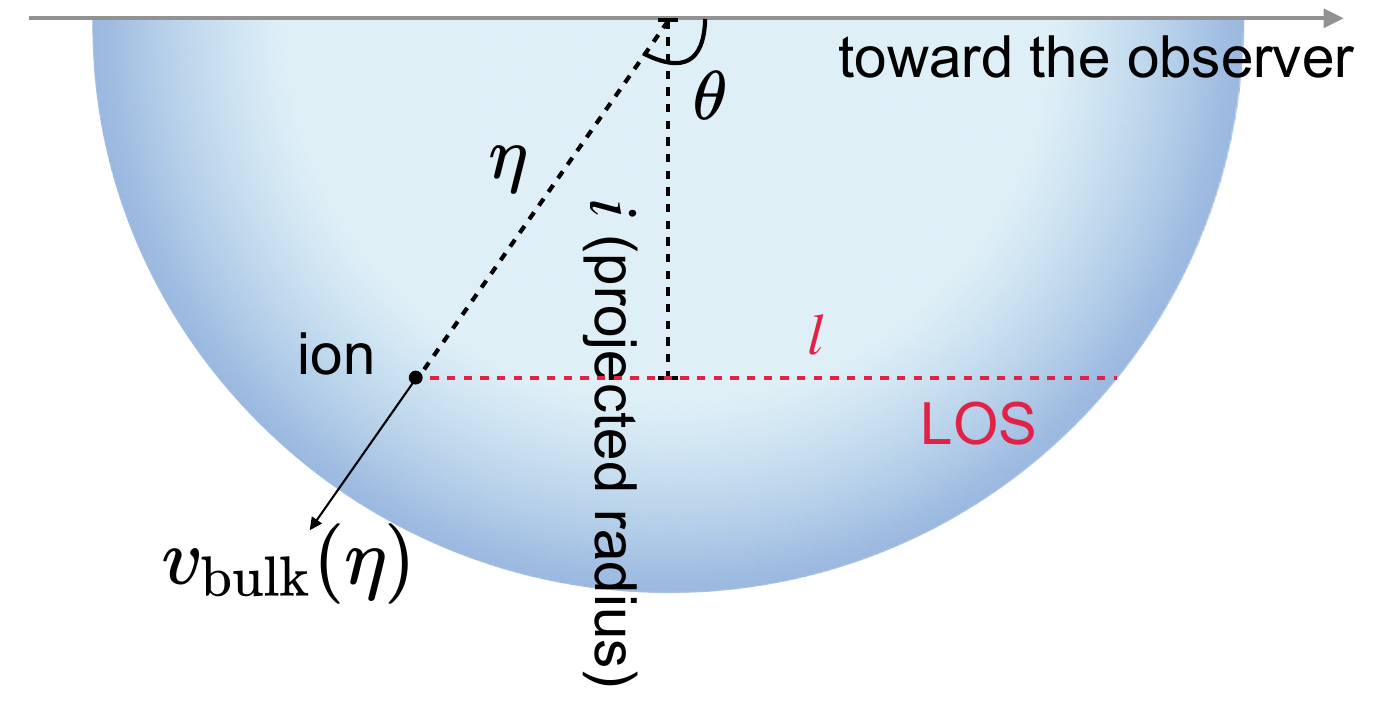}
}
\caption{Sketch for illustrating the line-of-sight optical depth. 
} 
\label{fig: photon_route}
\end{figure*}

We will characterize the importance of the RS effect in the SNR with
the optical depth along the line of sight (LOS) defined as
\begin{equation}\label{eq: tau}
    \tau_{\nu}(i)=R_s\int n_{\rm ion}(\eta,Z)s(\eta,x)dl,
\end{equation}
where $i$ and $l$ are the dimensionless projected radius and locational depth (both in units of $R_s$), respectively. As geometrically sketched in Figure\,\ref{fig: photon_route}, one have $\eta=[(\sqrt{1-i^2}-l)^2+i^2]^{1/2}$ and $0\leq l\leq 2\sqrt{1-i^2}$.

In the calculation of the LOS optical depth (Eq.\ref{eq: tau}), the cross section $s(\eta, x)$ should be obtained in the {\it observer's reference frame}.
In this frame, the expanding bulk motion, which can be resolved into 
the component along the LOS and transverse component vertical to the LOS, should be taken into account.
The dimensionless frequency shift $x(\eta)$ in Eq.\ref{eq: cross section} is here replaced with $x(\eta,\theta)=\{\nu-\nu_0 [1+\beta_{\rm bulk}(\eta)\cos\theta]\}/\Delta \nu_D(\eta,\theta)$, where $\beta_{\rm bulk}(\eta)=v_{\rm bulk}(\eta)/c$
and
$\cos\theta=(\sqrt{1-i^2}-l)/\{[(\sqrt{1-i^2}-l)^2+i^2]^{1/2}\}$.
Meanwhile, the Doppler width $\Delta\nu_D(\eta)$ in Eq.\ref{eq: cross section} is here replaced with $\Delta\nu_D(\eta,\theta)=[1+\beta_{\rm bulk}(\eta)\cos\theta]\nu_0(\sqrt{2}\sigma_v/c)$. 

\subsection{The Process and Parameters of the MC Simulations}
\subsubsection{Process of the simulations}
We apply MC simulations of RS effect to SNRs with gas distribution following the ST model (\S\ref{sec: ST Distribution}).
In the simulations, we use an algorithm adapted from the scheme developed in \citet{2018ApJ...861..138C}. 





\begin{figure}[ht!]
\centerline{
\includegraphics[scale=0.6]{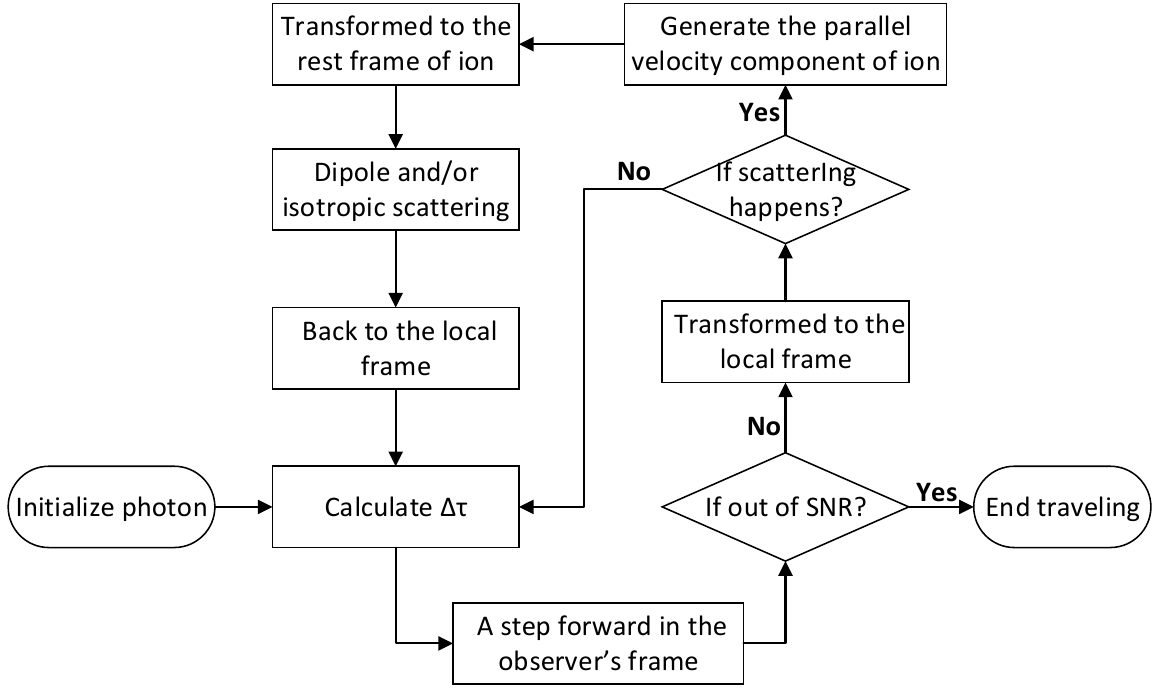}
}
\caption{The flow diagram of the simulation.} 
\label{fig: frame}
\end{figure}

A complete journey of photons simulated by the code is illustrated in Figure\,\ref{fig: frame}. More details of the process of the simulation are described below.

\textit{Stage 1.---}
We first give the information of the seed photons: 
photon's location, emitting direction, and energy. 
The seed photon's location is given in the observer's reference frame. Its positional angle of the photon is random, and its radius $r$ (or $\eta$) is obtained by means of the volume emission coefficient distribution of the plasma in the SNR. Specifically, $\eta$ is found from the probability distribution function \citep{2006ApJ...649...14D, 2018ApJ...861..138C}
\begin{equation}
 R_1=\frac{\int_0^{\eta} \nel(y)\nH(y)
 \epsilon(T(y)) y^2 dy}{\int_0^1 \nel(y) \nH(y) \epsilon(T(y)) y^2 dy}
    =\frac{\int_0^{\eta} G(y)dy}{\int_0^1G(y)dy}, 
\end{equation}
where 
$R_1\in[0,1]$ is a random number,
$\epsilon(T)=\zeta(Z) A(Z) f_{\rm ion}(T) f_{\rm u} (T)A_{\rm ul}$ 
is the emissivity of the line as a function of temperature and hence also of the location (with $f_{\rm u}(T)$ the ratio of the level population to the ion population), 
and $G(y)=\mathcal{G}^2(y)\epsilon(T(y))\,y^2$. 
In the calculation, the values of emissivity are adapted from the atomic database corresponding to temperature and ionization parameter (as visualized in Figure\,\ref{fig:em_OVII}).
The emitting direction of each seed photon is randomly generated in the local reference frame.
The energy of the seed photon is obtained also in the local frame according to the frequency distribution profile given by Eq.\ref{eq: line profile}.

\textit{Stage 2.---} 
We trace the photon after it is emitted or scattered at each step of the journey, with the stepsize in terms of optical depth 
\begin{equation}
\Delta \tau(\Delta\ell)=s(\eta,x)n_{\rm ion}(\eta,Z)\Delta\ell,  
\end{equation}
in which the cross section $s(\eta, x)$ is calculated in the local frame, and the adaptive spatial length step $\Delta\ell$ is got from the given $\Delta \tau(\Delta\ell)$.
With a sufficiently small $\Delta \tau(\Delta\ell)$, the plasma can be assumed to be uniform in each step.
In practice, we adopt $\Delta \tau(\Delta\ell)=0.001$, but set an upper limit for the spatial stepsize $\Delta \ell_{\rm max}=0.01\Rs$ to avoid a large spatial span.

\textit{Stage 3.---} 
It will be checked whether a scattering happens after the photon moves a spatial stepsize.
Before this step begins, the photon's moving direction and energy have been transformed from the local frame to the observer's frame.
A scattering event is judged to happen if the scattering probability $1-\exp(-\Delta\tau(\Delta\ell))$ is larger than a random number $R_2\in[0,1]$.
Then, the location of the scattered photon is recorded in the observer's frame.

To calculate the changes of the photon's energy and reemitted direction, the both quantities of the incident photon are transformed back from the observer's frame to the local frame.
We derive the component of the ion's velocity in the local frame parallel to the direction of the incident photon, $u_\|$ (in units of $\sqrt{2}\sigma_v)$, randomly from the RS probability that is a function of $u_\|$ and $x_i$ \citep{2002ApJ...578...33Z}
\begin{equation}
f(u_\|)=\frac{a(\eta)}{\pi}
\frac{e^{-u_\|^2}}{(x_i-u_\|)^2+a^2(\eta)}
H^{-1}(a(\eta),x_i).
\end{equation}
The two perpendicular components of the ion's velocity are given from Gaussian distribution with zero mean and standard deviation $\sigma_v$.

We transform the relevant quantities of the photon by coordinate rotation and Lorentz transformation between the local frame and the rest frame of the ion, in which the symmetric axis is along the incident direction.

We treat the remission of the photon as a combination of isotropic scattering (with a weight $w_{\rm I}$) and dipole scattering (with a weight $w_{\rm D}$), with a weighted probability distribution 
\begin{equation}\label{eq:R3}
R_3=\int_{-1}^{\mu}\left[\frac{1}{2}w_{\rm I}+\frac{3}{8}\lt(1+\mu'^2\rt)w_{\rm D}\right]\,d\mu',
\end{equation}
where $R_3\in[0,1]$ is a random number and $\mu$ is the cosine of the angle between the incident and reemitted directions of the photon. The values of $w_{\rm I}$ and $w_{\rm D}$ are given in Table\,1 in \citet{2018ApJ...861..138C} for oxygen lines O\,VIII Ly$\alpha$, O\,VII He$\beta$, and O\,VII He$\alpha$\,r.
With $\mu$ derived from Eq\ref{eq:R3}, we can obtain the reemission direction in the local frame.

If the scattering does not happen, the photon moves a step forward in the original direction.

\textit{Stage 4.---}
Once it is found to be outside the SNR boundary after a movement, the photon is thought to have escaped from the SNR, and its information after the last scattering is used as the final physical status of the photon.
 
In each simulation, $10^7$ photons are used as a compromise between good-quality statistics and economical calculations.

\subsubsection{Parameters used in the simulations}
We focus our simulations on the O~VII~He$\alpha$~r line. 
The atomic database AtomDB (v3.0.9) \citep{2012ApJ...756..128F} is used throughout this work, which provides the ion fraction, emissivity, atomic weight, the spontaneous emission rate for a temperature range up to $10^9$\,K. 
The gas temperatures higher than $10^9$\,K in the modeling are all replaced with this value.

\begin{figure}[ht!]
\centering
    \includegraphics[width=\textwidth]{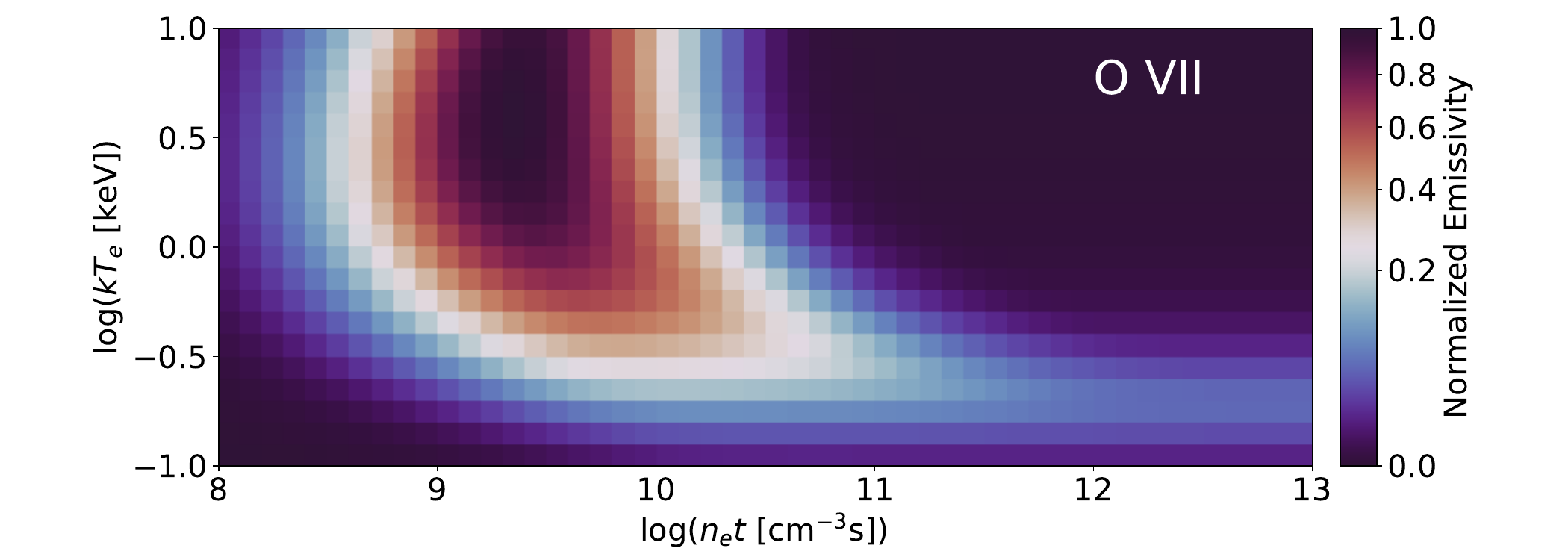}
\caption{Normalized emissivity of the \OVIIr\ line.}
\label{fig:em_OVII}
\end{figure}

For the shocked gas in SNRs with an ionization parameters $\nel\ti$ around $\sim10^{11}\cm^{-3}$\,s, the emissivity of the O\,VII\,r line peaks at temperature $\sim0.3 \keV$ (see Figure\,\ref{fig:em_OVII}). By comparison, the emissiviy is below half of the maximum at a temperature outside the range $\sim0.2$--$0.8\keV$.
In view of such physical conditions, Cygnus Loop appears to be a very suitable target to examine the O\,VII\,r emission.

Table\,\ref{tab:simulation parameter} gives basic parameters (radius, shock velocity, ionization timescale, explosion energy, preshock gas density, and Sedov age) of an SNR, exemplified by Cygnus Loop, for our simulations.
\citet{2003ApJ...584..770R} analyzed the FUSE and ROSAT data and estimated a shock speed $\Vs$ of $350\,\km \ps$ by the northeast H$\alpha$ profile. \citet{2023ApJ...948...97S} obtained a shock velocity of $180\km\ps$ and $240\km\ps$ by the H$\alpha$ and [O~III] $\lambda\,5007$ emission. \citet{2009ApJ...705.1152U} conducted a comprehensive study on the shell structure in the northeast and southwest of the remnant (with a mean radius 13\,pc), and the derived values of the shock velocity vary in the range 350--$450\km\ps$. The X-ray emission near the southwestern rim is fitted with plasma temperatures in the range 0.15--0.20 keV \citep{2013ApJ...764...55L}, corresponding to shock velocities $\sim350$--$400\km\ps$. 

The numerical study by \citet{2011A&A...527A..55P} implicated a low value of the explosion energy of 6--8$\times10^{49}$ erg, and recent modelling by \citet{2017MNRAS.464..940F} also give a subenergetic explosion energy of 2$\times10^{50}$ erg.
The typical values of the parameters are adopted and divided into two cases.
In the both cases, the explosion energy and preshock density satisfies $E/n_0=(25/4)(1.4\mH/\xi)\Rs^3\Vs^2$.

To simplify the model, 
we use a single ionization timescale $n_{\rm e}\ti$ : we obtain the $1.2n_{\rm H}(\eta)\,\ti$ value at the
half-mass radius, and assign it to all the internal hot gas.
Here, technically, we also approximate $\ti$ at radius $\eta$ or $r$ to be the time duration it takes the shock front to move from $r$ to $\Rs$. 
In the concerned temperature range $\sim 0.1$--$0.4$\,keV, the emissivity does not vary drastically with $\nel\ti$ within $\sim 10^{10}$--$10^{11}\,\cm^{-3}\s$ (see, e.g., Figure\,\ref{fig:em_OVII}).
The $n_{\rm e}\ti$ values thus adopted in Table\,\ref{tab:simulation parameter} for the two cases ($\sim5\E{10}\cm^3$\,s) are similar to those obtained for the knot at the southwestern shell (``SW-K") of Cygnus Loop from the {\it XMM-Newton} spectra
\citep{2019ApJ...871..234U}.
The single $\nel\ti$ assumed for the gas is a good approximation that can be simply tested by comparison with the case of $\nel\ti$ varing with radius.

\begin{figure}[ht!]
\centerline{
\subfigure[]{
        \includegraphics[scale=0.6]{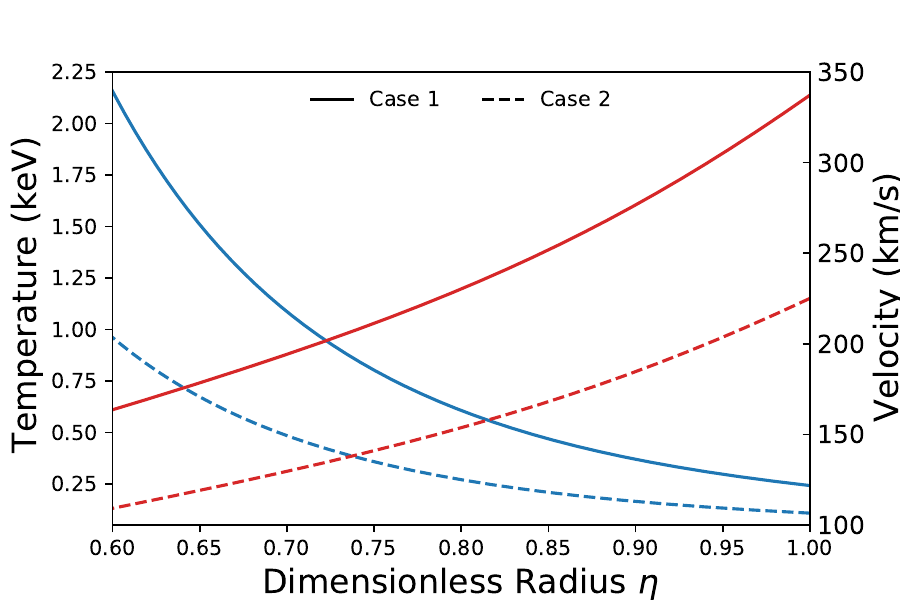}
        \label{fig: distribution}
    }
    \subfigure[]{
        \includegraphics[scale=0.6]{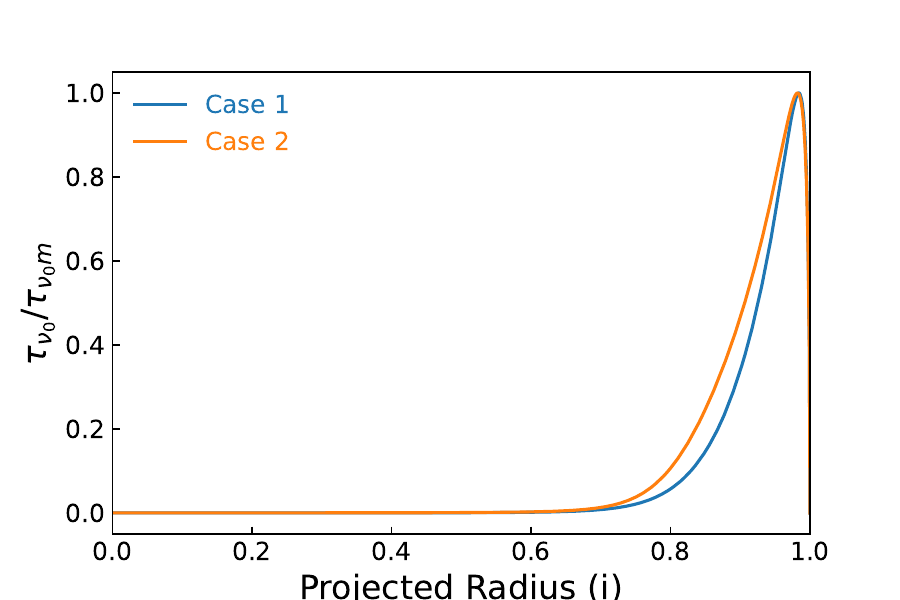}
        \label{fig: tau_case}
    }
}
\caption{(a): The ST-model temperature (blue lines) and bulk motion velocity (red lines) distribution of the hot gas.
(b): The distribution of the line-center optical depth of O~VII~He$\alpha$~r line with the projected radius for Case~1 and Case~2. The values of optical depth are normalized by the maximum value $\taum$
in each case.}
\label{fig: model}
\end{figure}

Figure\,\ref{fig: distribution} presents the temperature and expansion velocity for the two cases in the Sedov solution, with the $\Vs$ values given in Table\,\ref{tab:simulation parameter} used.
Because the ions distribute mainly in the outer region (i.e., $\mathcal{G}$ is very small for $\eta<0.6$ in Figure\,\ref{fig: STsolve})
, we only plot the distribution of the outer region between $\eta=0.6$ and 1.
Incorporating the distributions of gas density, temperature, and bulk motion velocity as well as the single ionization timescale, 
the line-center optical depth of O\,VII He$\alpha$\,r line  $\tau_{\nu_0}$ as a function of the projected radius, normalized by the maximum value
, is calculated \inlet{using} Eq.(\ref{eq: tau}) and is presented in Figure\,\ref{fig: tau_case}.
This maximum value of $\tau_{\nu_0}$ is denoted as $\taum$.
As seen in Figure\,\ref{fig: tau_case}, $\tau_{\nu_0}$ peaks at $i=0.98$, very close to the visual boundary. 
We note that, although $\taum$ varies with the elemental abundance ($\taum\propto \zeta$), the $\tau_{\nu_0}/\taum$ distribution is irrespective of $\zeta$, as long as other parameters keep unchanged.

Particularly, for O~VII~He$\alpha$~r line, the maximum $\tau_{\nu_0m}$ reaches $1$%
, which indicates the gas becoming optically thick, when $\zeta({\rm O})$ is ${\sim}0.3$ in Case~1 or $\sim0.05$ in Case~2. 

    

\begin{table}[!ht]
    \centering
    \caption{Parameter values used in simulations}
    \begin{threeparttable}
    \begin{tabular}{*8{c}}
    \toprule
         & $\Rs$ & $\Vs$  & $E$ & $n_0$  & $n_e\ti$\tnote{a} & $t$  &$T_{\rm s}$\\
         & (pc) & ($\km\ps$)  & ($10^{51}$ erg) & ($\cm^{-3}$) & ($10^{10}{\cm^{-3}}{\rm s}$)& ($10^4$\,yr) & (keV)\\ 
    \midrule 
Case 1 & 13  & 450 & 0.33 & 0.35 & $\sim$4.8 & $\sim 1.1$ & $\sim0.24$  \\
Case 2 & 13  & 300  & 0.42 & 1.0 & $\sim20$  & $\sim1.7$ & $\sim0.11$\\
    \bottomrule
    \end{tabular}
    \tablecomments{The ST model parameters used in our simulations. }
    \begin{tablenotes}
    \footnotesize
    \item[a] The values of $\nel\ti$ here are assumed for the gas at the 
    half-mass radius.
    \end{tablenotes}
    \end{threeparttable}
    \label{tab:simulation parameter}
\end{table}

\section{Results} \label{sec: Results}
\begin{figure}[ht!]
\centering
\subfigure[for entire projected region]{
    \includegraphics[scale=0.6]{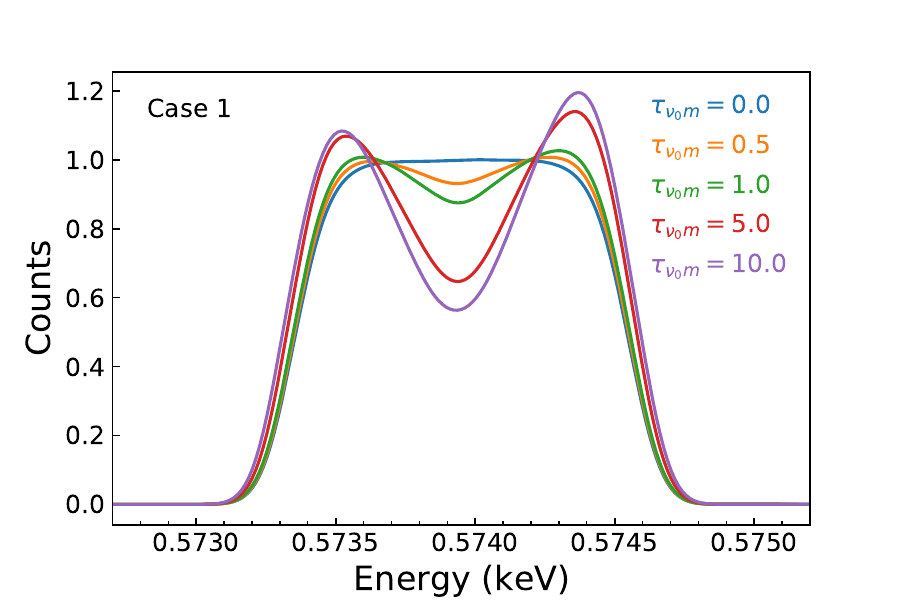}
    \includegraphics[scale=0.6]{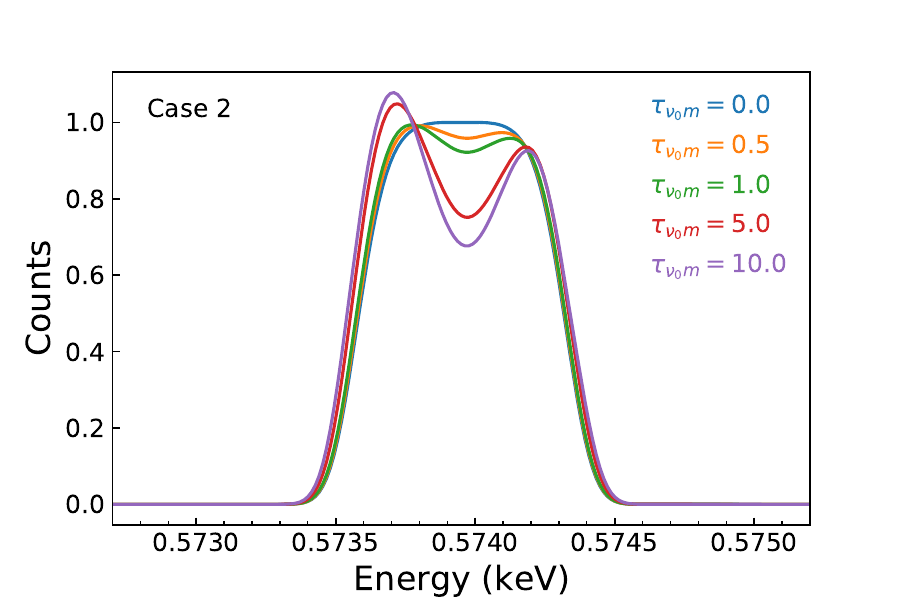}
    \label{fig: line_all}
}
\subfigure[for inner projected region]{
    \includegraphics[scale=0.6]{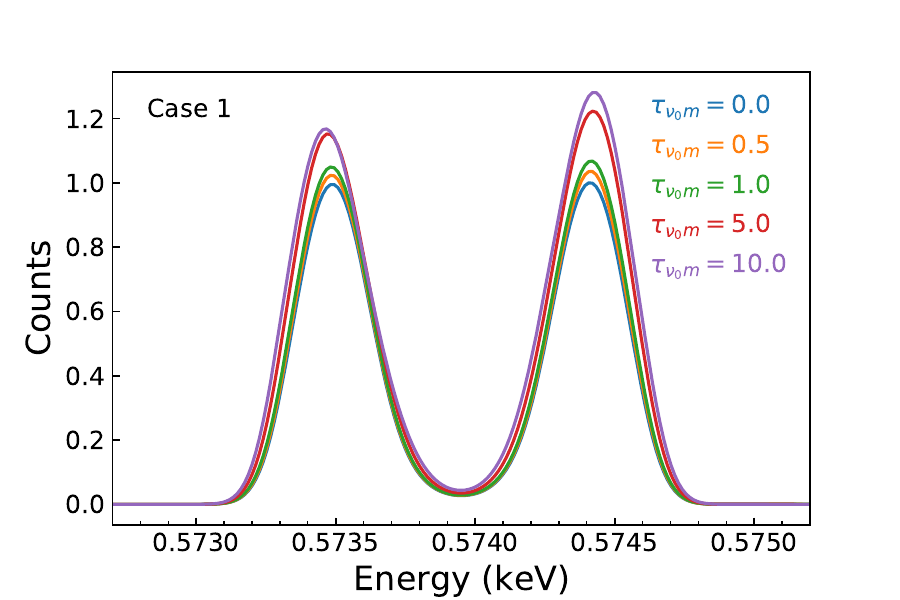}
    \includegraphics[scale=0.6]{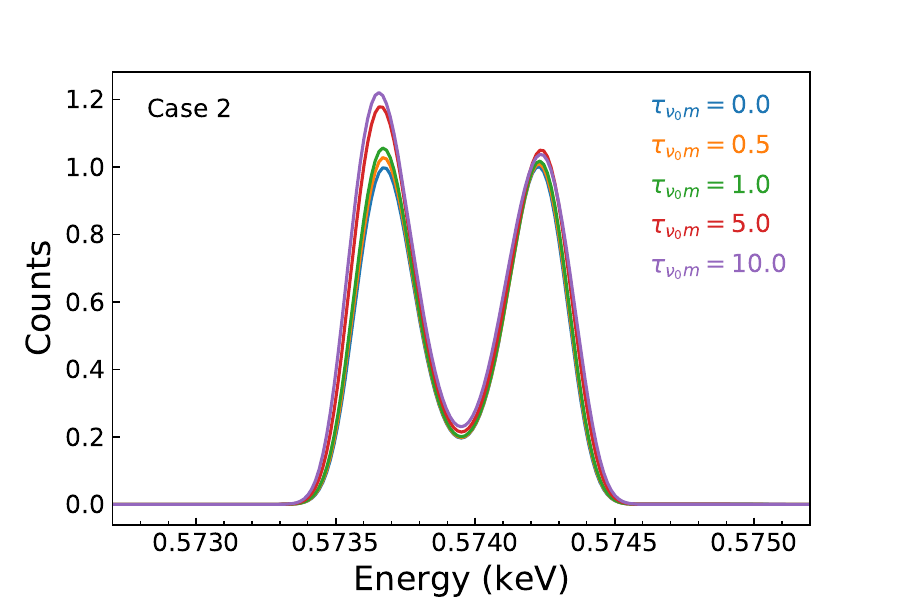}
    \label{fig: line_inner}
}
\subfigure[for outer projected region]{
    \includegraphics[scale=0.6]{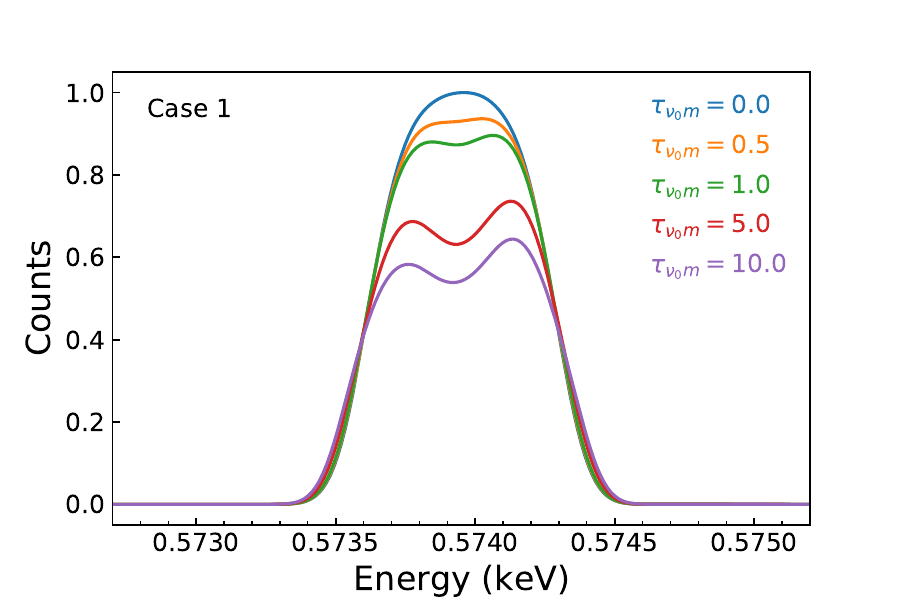}
    \includegraphics[scale=0.6]{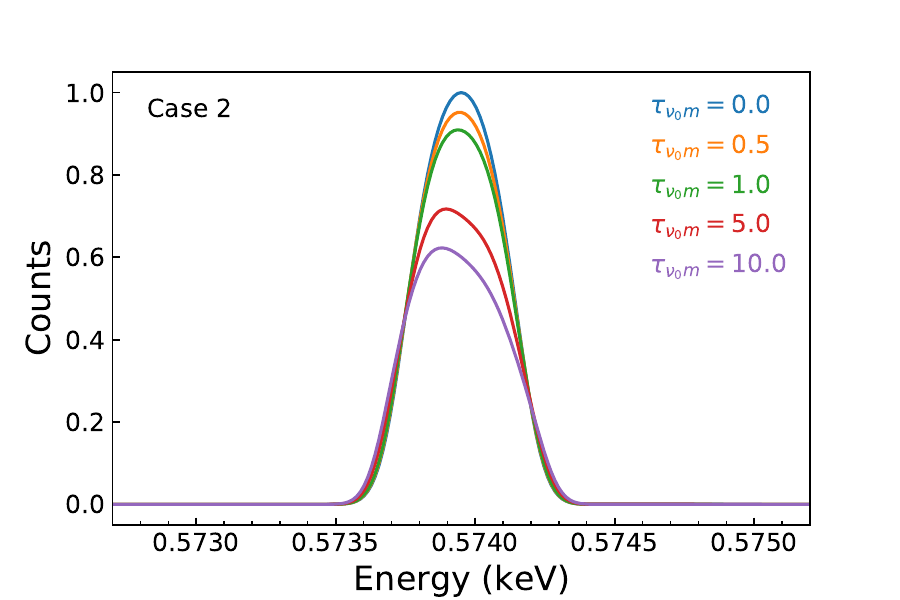}
    \label{fig: line_outer}
}
\caption{Line profiles of \OVIIr\ photons received by the observers. (a) For the photons from the entire SNR; (b) for the photons from the inner region (with projected radius $i< 0.8$); (c) Fot the photons from the outer region (for projected radius $i>0.8$). A Gaussian smoothing with energy interval of 0.04\,eV has been applied to reduce the noise of the simulated data.} 
\label{fig: lineprofile}
\end{figure}

\subsection{Line Profiles}\label{sec:lineprofile}
Figure\,\ref{fig: lineprofile}
presents the line profiles of \OVIIr\ photons observed from three different (entire, outer, and inner) regions of the SNR for various optical depths for the two cases described in Table\,\ref{tab:simulation parameter}. 
The panels in the left column are for Case~1, and those in the right column are for Case~2.
For the entire SNR (see Figure\,\ref{fig: line_all}, i.e.\ the two top panels),
the line profiles without scattering ($\taum=0$) appear flat and broad at the top (compared with a Gaussian) due to Doppler broadening of the gas's bulk motion.
As $\tau_{\nu_0m}$ increases, the line profiles get increasingly deformed: 
the photon counts at the line core are increasingly reduced, and double-peak or 
saddle-like patterns appear for large $\taum$,
but the total counts do not change.
The line broadening resulting from the scattering effect is slight compared with the Doppler broadening.

For the inner region ($i<0.8$; see Figure\,\ref{fig: line_inner}, i.e.\ the two middle panels), the profiles are separated into two components due to Doppler effect. As $\tau_{\nu_0m}$ increases, the two components are slightly broadened and rise monotonously, resulting from the spatial redistribution of the resonantly scattered photons. 

For the outer region ($i>0.8$; see Figure\,\ref{fig: line_outer}, i.e.\ the two bottom panels), the photon counts are diminished with the increase of $\tau_{\nu_0m}$.
The diminished photons are redistributed via RS to the inner region.
Since the gas in the outer region moves with a relatively low velocity component along the LOS and suffers lighter Doppler effect, the line profiles are not separated as those of the inner region.

Additionally, it is noted that the double-peak patterns deviate from bilateral symmetry for large $\taum$. In Case~1, the redward-side peak is lower than the blueward-side one, while in Case~2, it is vice versa (such phenomena will be qualitatively explained in \S\ref{sec:Asymmetry of Line Profiles}).

\subsection{The Radial Distribution of Surface Brightness}
\begin{figure}[ht!]
\centerline{
\includegraphics[scale=0.5]{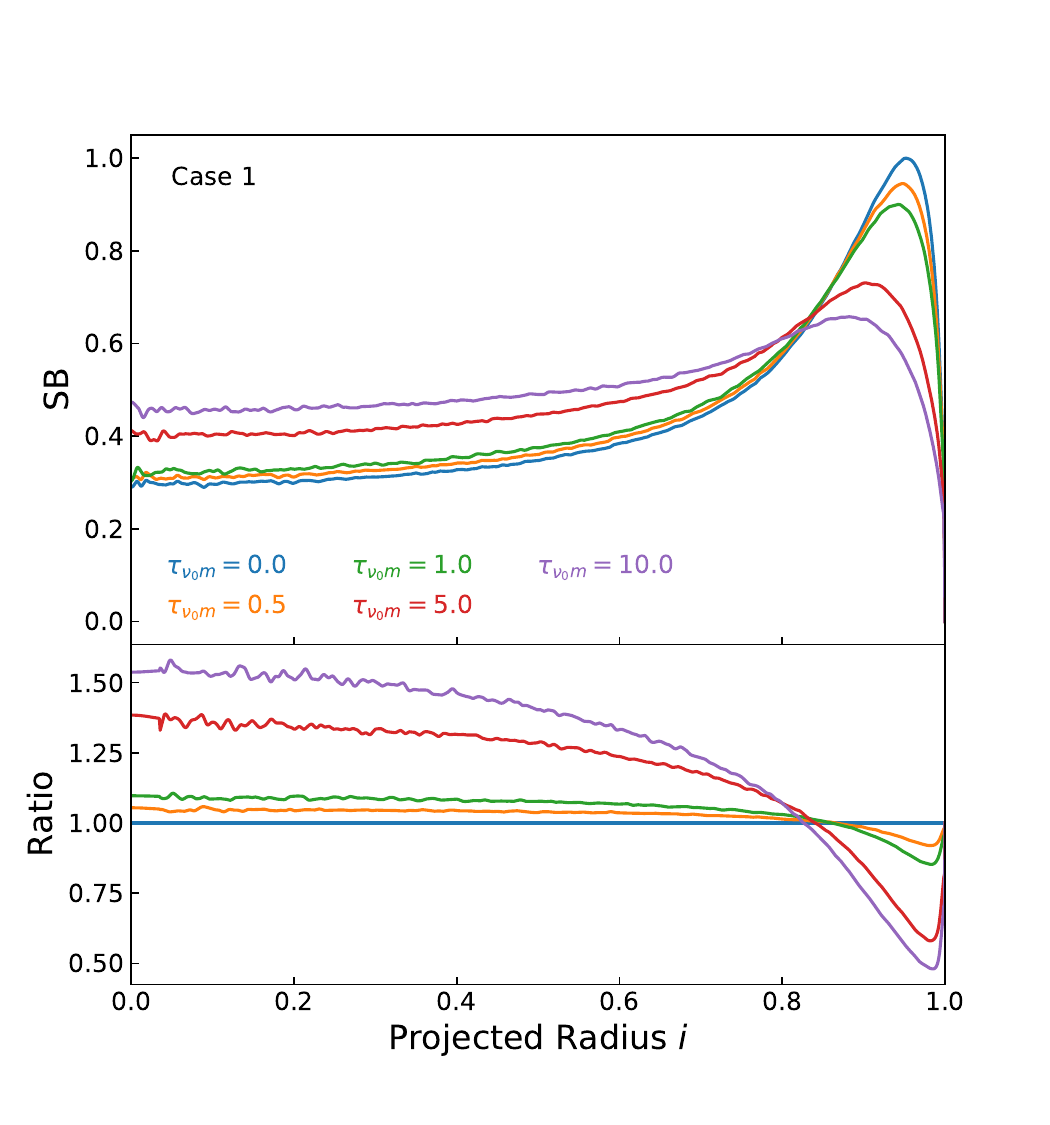}
\includegraphics[scale=0.5]{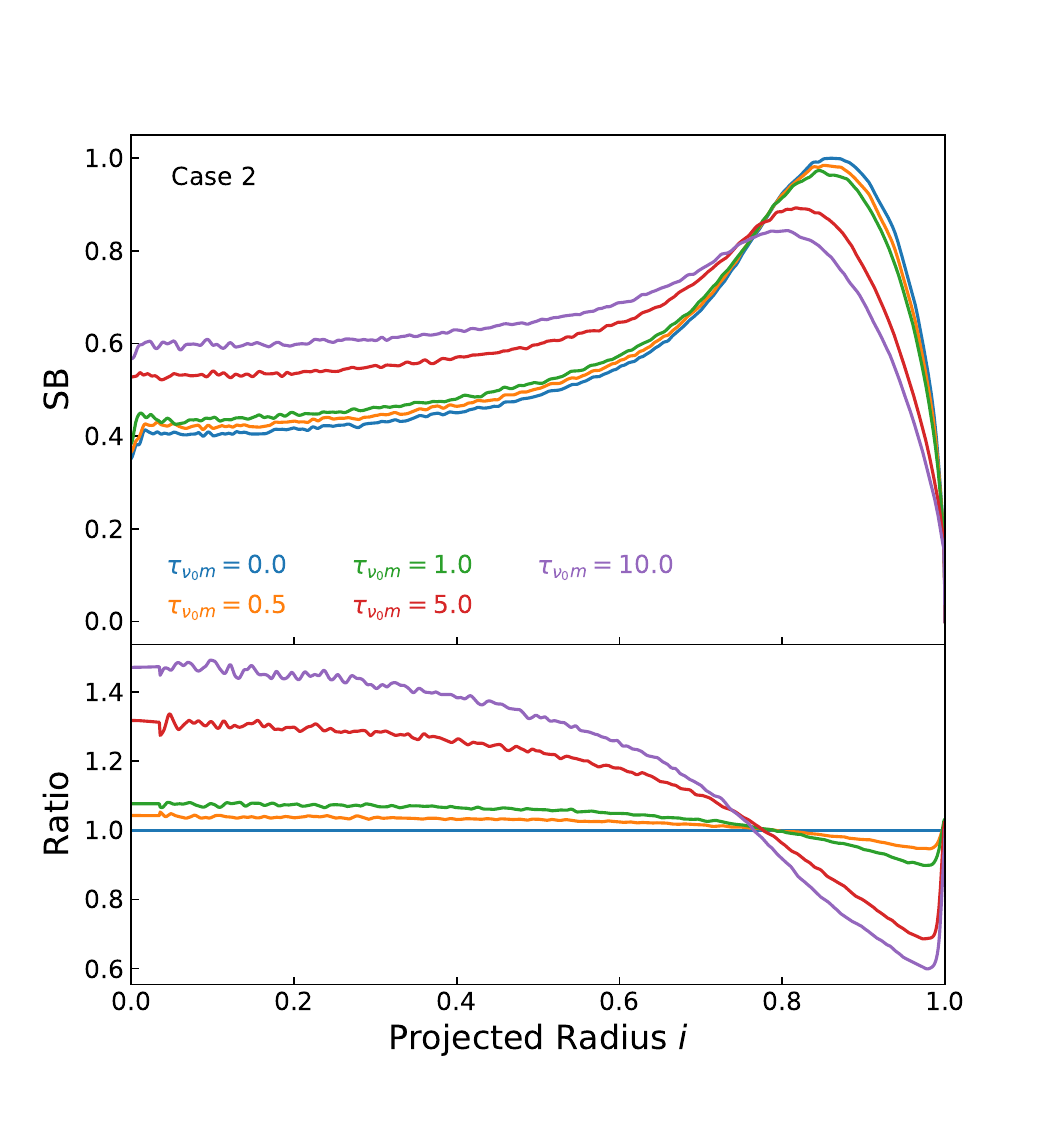}
}
\caption{Top: Projected radial SB distribution of O~VII~He$\alpha$~r line for Case~1 and Case~2. 
The plotted SB values are scaled with the maximum SB value for $\tau_{\nu_0m}=0$. 
Bottom: Ratio of the projected radial SB to the SB without RS effect (i.e., for $\tau_{\nu_0m}=0$).}
\label{fig: SB}
\end{figure}

Figure\,\ref{fig: SB} shows the distribution of projected radius surface brightness (SB) profiles of \OVIIr\ line emission for Case~1 and Case~2. 
The SB of the emission peaks near the edge of SNR for various $\tau_{\nu_0m}$ values. 
With the increase of $\taum$, the brightness of the outer region (essentially for $i\ga0.8$
) drops, the brightest location gets inward, and the inner region becomes brightened. 
This trend is consistent with the distortion of the line profiles of the inner and outer regions (\S\ref{sec:lineprofile}
Certainly, the deformed SB profiles are caused by the spatial redistribution of the scattered photons.


\subsection{Line Ratios}
\begin{figure}[ht!]
\centerline{
    \includegraphics[scale=0.6]{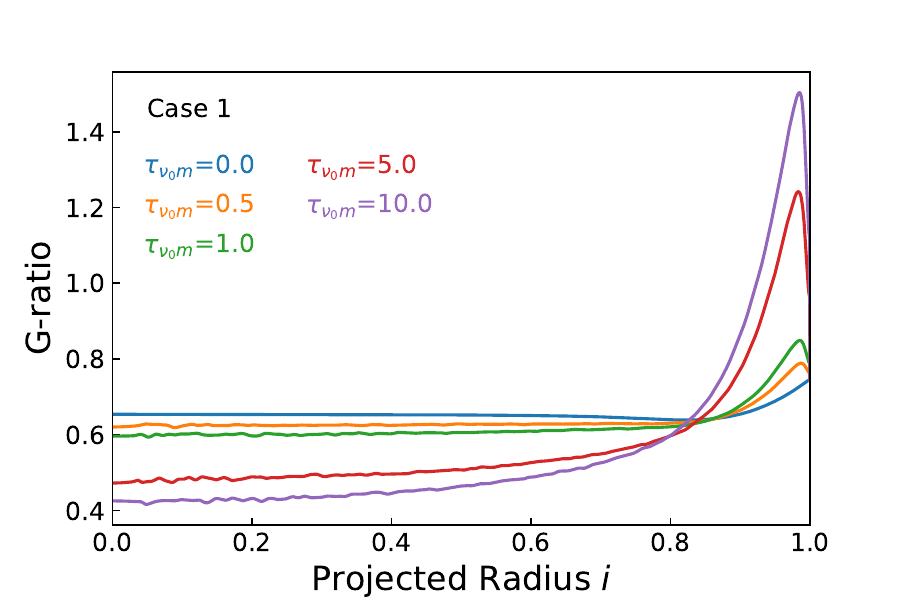}
    \includegraphics[scale=0.6]{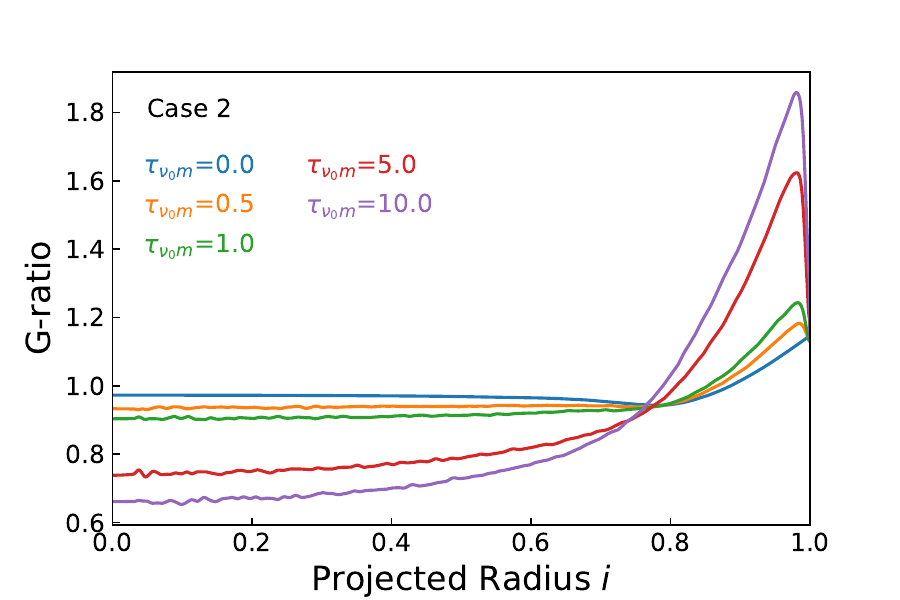}
}
\caption{Projected radial profiles of the O\,VII He$\alpha$ triplet G-ratio for various $\taum$ values for both Case~1 and Case~2. 
}
\label{fig: Gratio}
\end{figure}

Figure\,\ref{fig: Gratio} presents the radial profiles of the O\,VII He$\alpha$ triplet G-ratio for various $\taum$ values.
In the calculation, we obtain the SB of the intercombination (i) line, forbidden (f) line, and the r line without scattering by integrating the volume emission coefficient along the LOS
at projected radius $i$:
\begin{equation}\label{eq:I_line)}
    I_{\rm line}(i)\propto
    \int \nel(\eta)\nH(\eta)\epsilon(T(\eta)) dl.
\end{equation}
Then the G-ratios are obtained by dividing 
the ratio $[I_{\rm i}(i)+I_{\rm f}(i)]/I_{{\rm r},\taum=0}$ by
the ratio between the SB of O\,VII\,r line for various $\taum$ and the SB without scattering (i.e., represented by the curves in the bottom panels of Figure\,\ref{fig: SB}).
As a consequence of the redistribution of SB due to RS effect, the 
the G-ratio in the two cases both raises evidently in the outer region ($i\ga0.8$) and drops in the inner region. 

\begin{figure}[ht!]
\centerline{
    \includegraphics[scale=0.6]{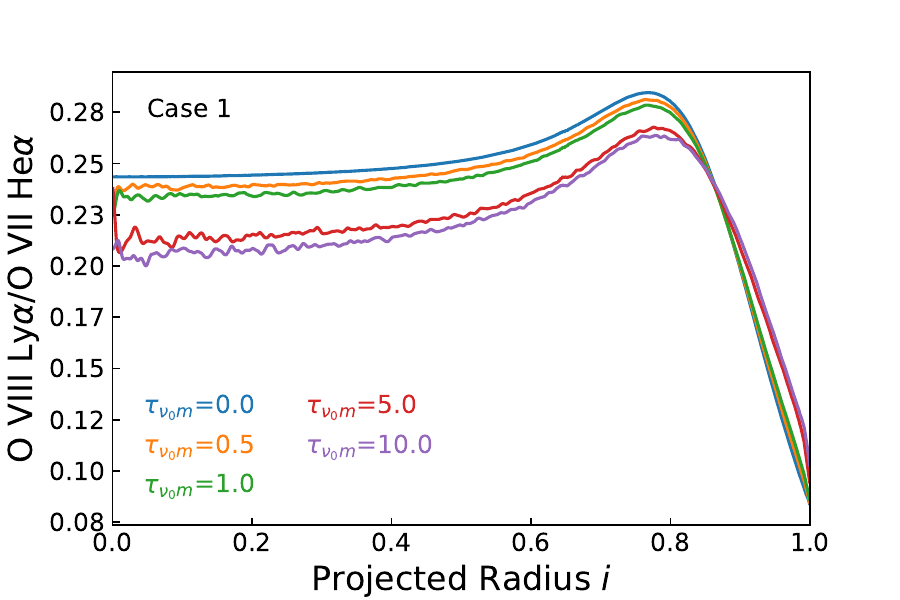}
    \includegraphics[scale=0.6]{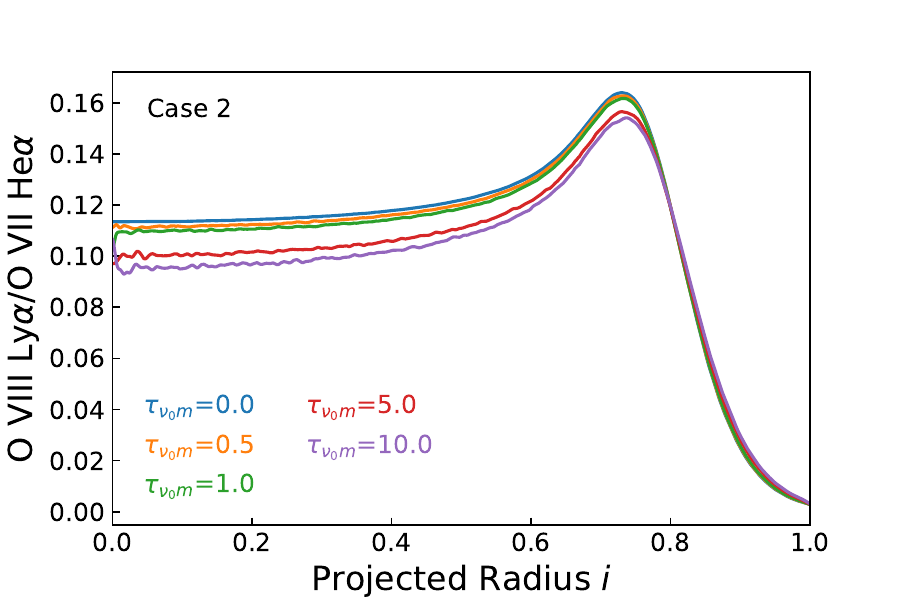}
}
\caption{Projected radial profiles of the OV\,III Ly$\alpha$/O\,VII He$\alpha$ ratio for various $\taum$ of the \OVIIr\ line for both Case~1 and Case~2. 
}
\label{fig: OVII_VIII_ratio}
\end{figure}

Our calculation is here extended from the \OVIIr\ line to the O\,III\,Ly$\alpha$ doublet \citep[see Table\,1 in][]{2018ApJ...861..138C}.
The two lines suffer from RS effect in different degrees. Therefore the O\,III~Ly$\alpha$/O\,VII~He$\alpha$ ratio can be changed, which may result in biased estimations on the temperature and the ionization parameter.
Figure\,\ref{fig: OVII_VIII_ratio} shows the surface radial distribution of the ratio between the O\,III\,Ly$\alpha$ doublet 
(which is optically thin here, with optical depths $<0.05$)
and the O\,VII\,He$\alpha$ 
triplet for various $\taum$ of the \OVIIr\ line.  
The ratio in the both cases peaks at around $i\sim0.8$.
In each case, we find the O\,III~Ly$\alpha$/O\,VII~He$\alpha$ ratio differs slightly for various $\taum$, which indicates RS has small impact to the line ratio and the inferred temperatures and ionization parameters. 

\section{Discussions}\label{sec: Discussions}

\subsection{Asymmetry of Line Profiles}\label{sec:Asymmetry of Line Profiles}
The asymmetric line profiles obtained in our simulation deviate from bilateral symmetry for large $\taum$, which is clearly different from the case of static gas.

\begin{figure}[ht!]
\centering
{
    \includegraphics[width=.5\textwidth]{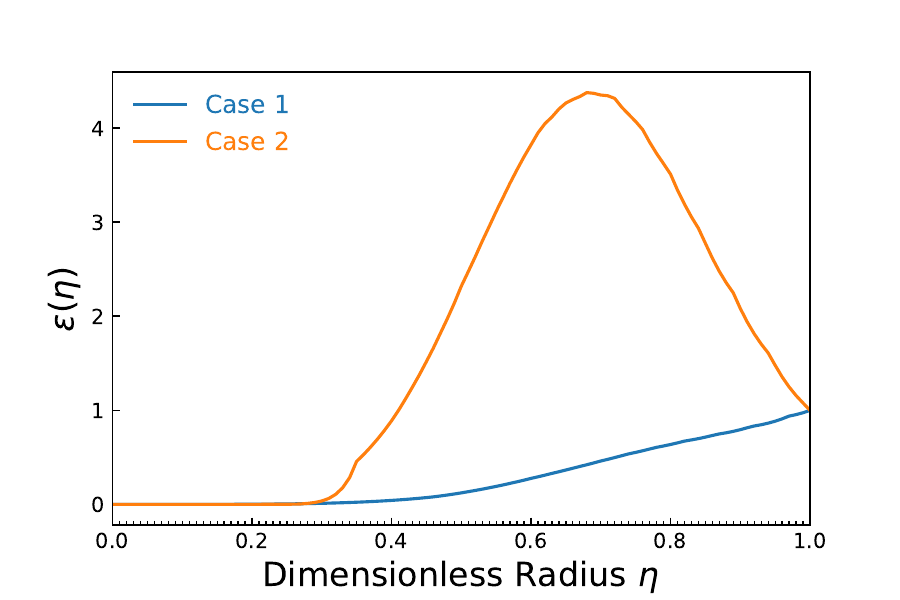}
}
\caption{The radial distribution of seed \OVIIr\ photon production efficiency per OVII ion, $\varepsilon$, normalized at $\eta=1$, for Case~1 and Case~2.}
\label{fig:varepsilon}
\end{figure}

The asymmetry of the line profile reflects the total effect of scattering of the photons emitted by the OVII ions in bulk motion at different radii with different emissivity.
It could be briefly explained in terms of the ratio (normalized at $\eta=1$) between the volume emission coefficient $\nel(\eta)\nH(\eta)\epsilon(T(\eta))$ and the O\,VII ion density $n_{\rm OVII}(\eta)$:
\begin{equation}
    \varepsilon=\epsilon(T(\eta))\mathcal{G}(\eta)/f_{\rm ion}(T(\eta)),
\end{equation}
which stands for an efficiency of the \OVIIr\ line emission from per OVII ion.
As seen in Figure\,\ref{fig:varepsilon}, the \OVIIr\ emitting efficiency $\varepsilon$ in Case~1 increases monotonically with radius $\eta$, indicating that the seed photon production is most efficient in the outermost part of the SNR shell.
In Case~2, $\varepsilon$ is not distributed monotonically and the photon production is less efficient in the outermost part than that in the inner part. 

\begin{figure}[ht!]
\centerline{
    \includegraphics[scale=0.6]{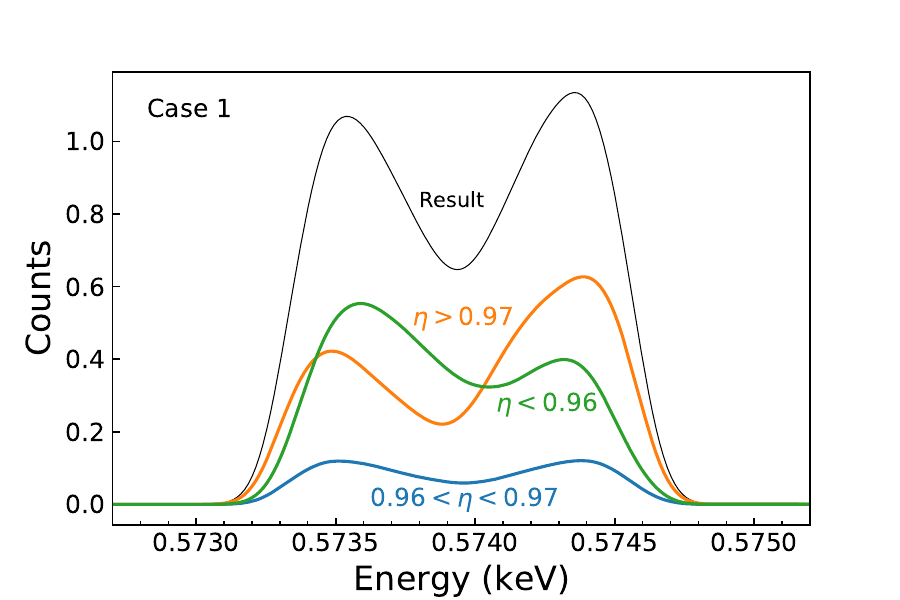}  
    \includegraphics[scale=0.6]{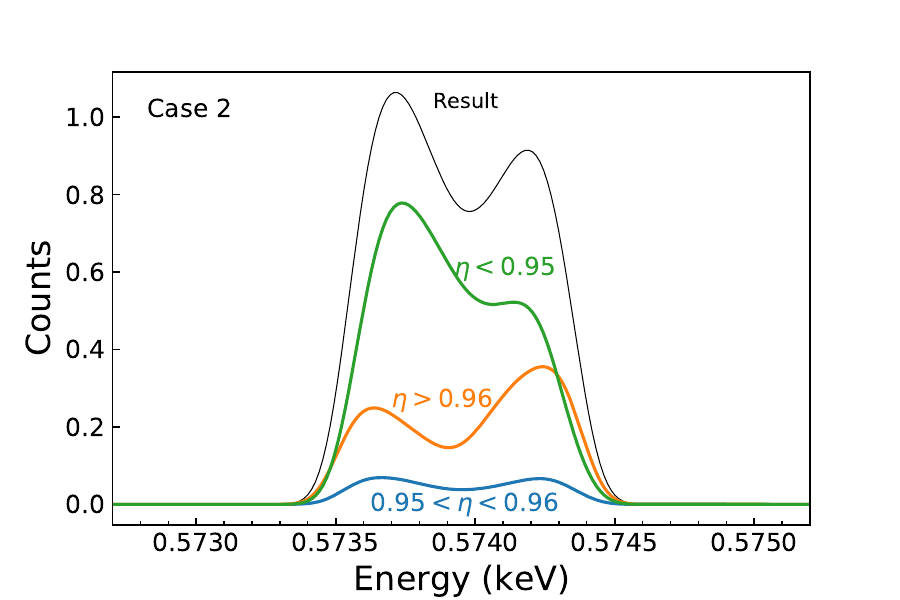}
    }
\caption{The line profiles of the scattered photons starting from different radial parts of the remnant.  $\taum=5$ is set for the both cases.
}
\label{fig:diff-profiles}
\end{figure}

For the photons produced in the outer part of the shell, those produced in the anterior half, which are majorly distributed in the blue side of the line profile, experience less scatterings than those that are produced in the posterior half but majorly scattered by the gas of the posterior half of the inner part of the shell
(because the cross section of scattering by the anterior gas is small due to large frequency shift $x$ caused by the Doppler effect), 
and therefore the line profile of the photons from the outer part of the shell is higher in the blue side than in the red side, as represented by the orange curves in Figure\,\ref{fig:diff-profiles}.
On the other hand, 
for the photons produced in the inner part of the shell, those produced in the posterior side, which are majorly distributed in the red side of the line profile, experience less scatterings than those that are produced in the anterior half, 
and therefore the line profile of the photons from this part of the shell is higher in the red side then in the blue side, as represented by the green curves in Figure\,\ref{fig:diff-profiles}.
We can find an intermediate layer in the shell, the photons produced in which are nearly equally distributed in the two sides of the line profile. For $\taum=5$, such a layer is in $0.96<\eta<0.97$ for Case~1 and in $0.95<\eta<0.96$ for Case~2.

In Case~1, the \OVIIr\ photon production is more efficient in the outer part than in the inner part, and thus, on the whole, the blue side is higher than the red side.
In Case~2, the photon production is more efficient in the inner part, and as a result, the red side dominates. 




\subsection{Line Profile Measurement}\label{sec: Measurement}
\begin{figure}[ht!]
\centering
\subfigure[for entire projected region]{
        \includegraphics[scale=0.6]{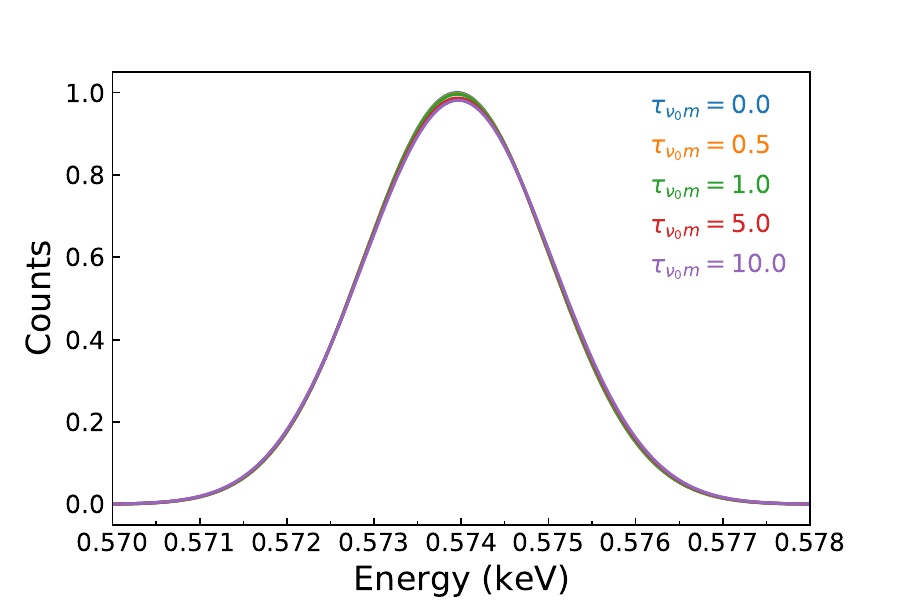}
        \includegraphics[scale=0.6]{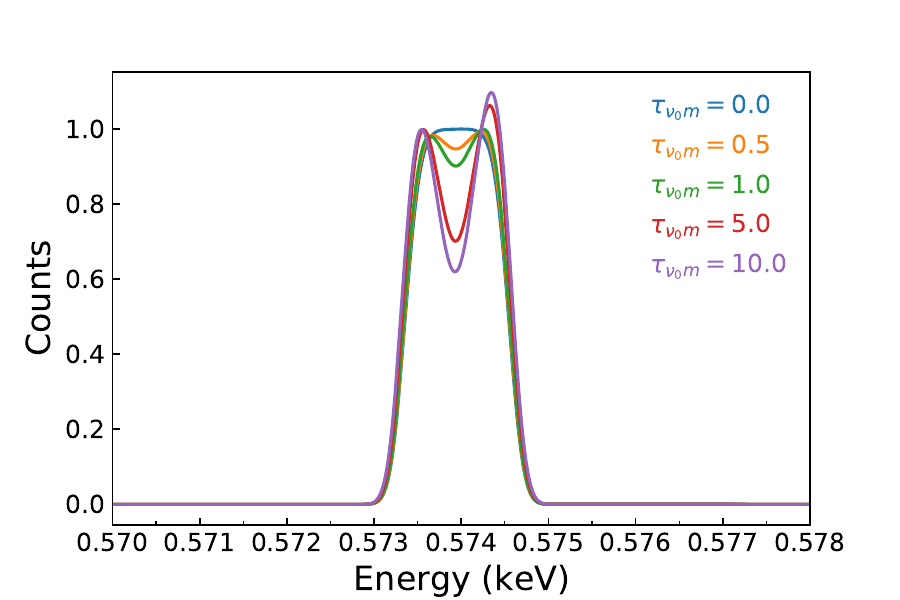}
    }
\subfigure[for inner projected region]{
        \includegraphics[scale=0.6]{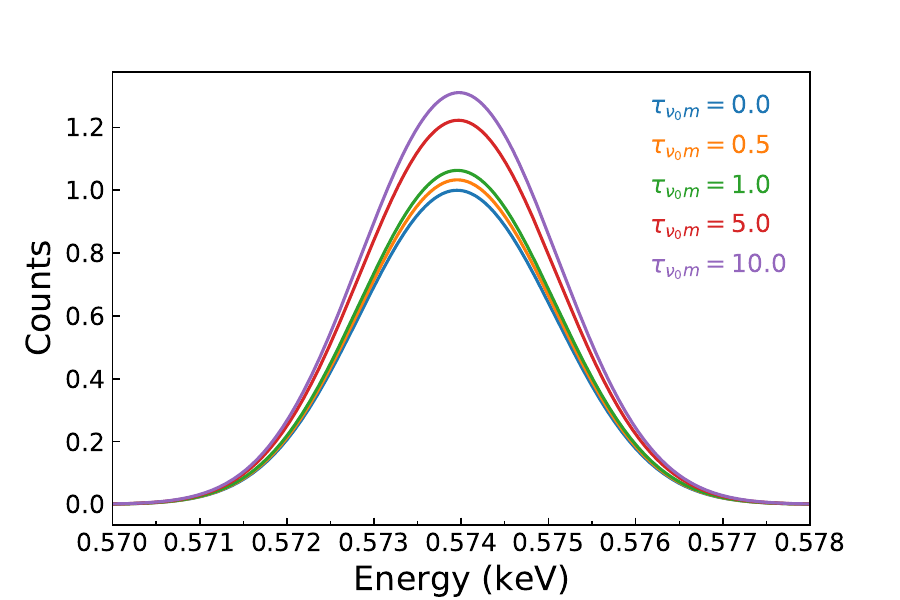}
        \includegraphics[scale=0.6]{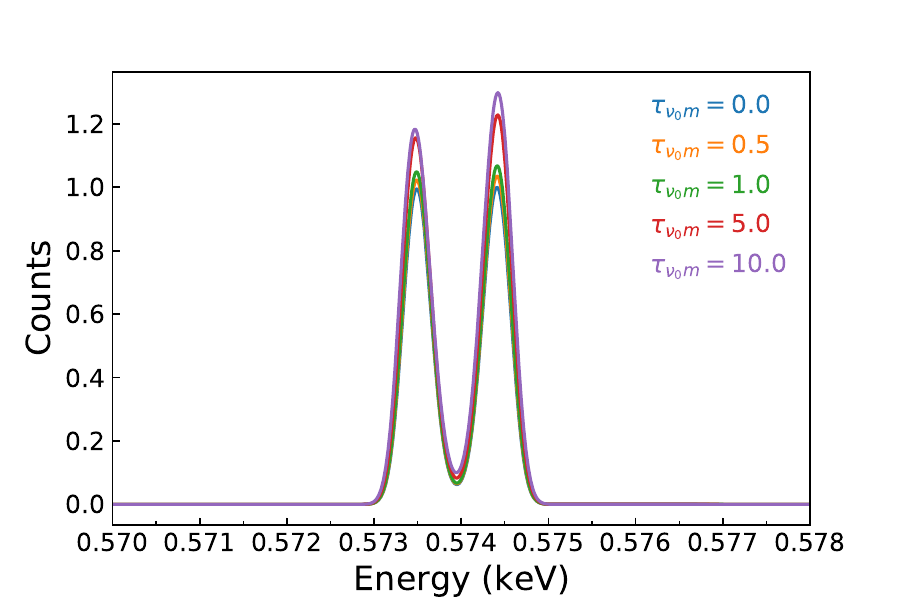}
    }
\subfigure[for outer projected region]{
        \includegraphics[scale=0.6]{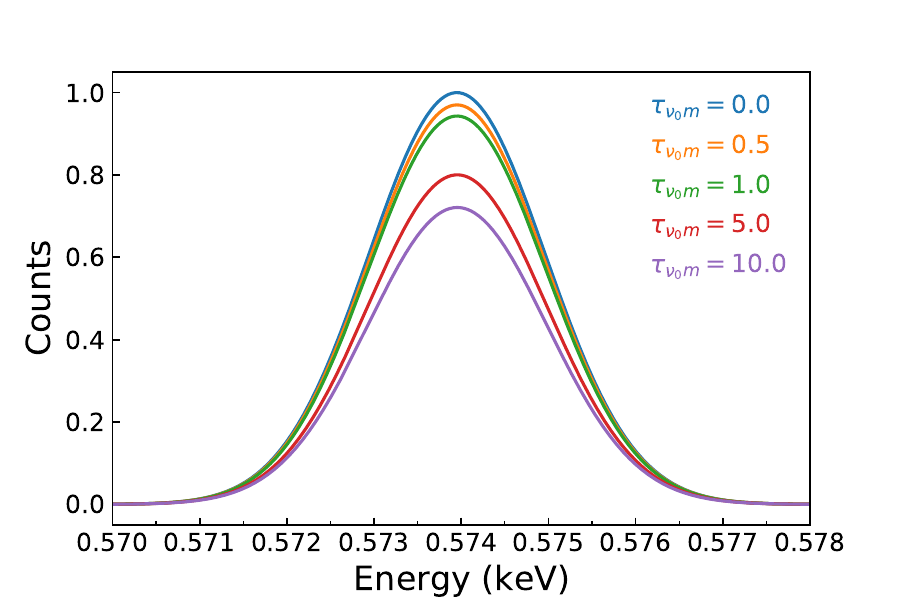}
        \includegraphics[scale=0.6]{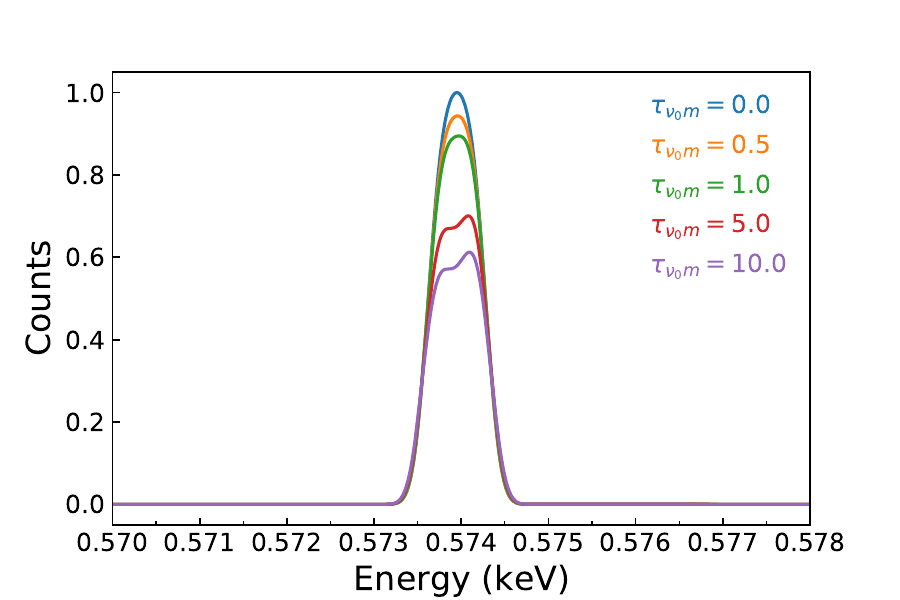}
    }
\caption{The line profile convolue with Gaussian of 1 eV (left column) and 0.1 eV (right column) energy resolution for various $\taum$. Here present the results for Case~1 only. }
\label{fig: lineprofile_conv}
\end{figure}
RS results in considerable deformation of line profiles (Figure\,\ref{fig: lineprofile}). 
However, the line widths are about 1\,eV, which cannot be resolved by X-ray observation with the state-of-the-art energy resolution ($\ga2$\,eV for O\,VII line).

We calculate the line profiles of Case~1 that are supposed to be observed with energy resolution 1\,eV and 0.1\,eV by convolving the line profiles in Figure\,\ref{fig: lineprofile} with two Gaussian profiles, respectively.
The left column of Figure\,\ref{fig: lineprofile_conv}  (convolved with 1\,eV) shows that the observed profiles are singly peaked, other than doubly peaked.
There are very small blueward shifts of line center of order 0.01\,eV, basically imperceptible. 
Meanwhile, the line profiles convolved with 0.1\,eV in the right column reserve most characteristics of the theoretical ones.  
For the inner and outer regions (see the middle and bottom rows in Figure\,\ref{fig: lineprofile_conv}), however, the flux change may be perceptible even for the 1\,eV convolution.


We note that 
some of next-generation X-ray observational missions will achieve an energy resolution of $\la1$\,eV. For example, HUBS will present a resolution at least down to 1--2\,eV, with a goal of 0.5\,eV, in soft X-ray band \citep{2020JLTP..199..502C},
and Arcus is designed to have an energy resolution of 0.16--0.2\,eV \citep{2022ApJ...934..171H}. Thus, the spectral features of the \OVIIr\ line subjected to RS could be expected to be observed before long.


\subsection{The RS of Cygnus Loop}\label{sec:The RS of Cygnus Loop}

Cygnus Loop (G74.0$-$8.5) is a nearby core-collapse shell-like SNR
at a distance of 540 pc \citep[e.g.,][]{2005AJ....129.2268B}. It is a 
middle-aged ($\sim$ 1--2$\times10^4$\,yr) remnant as estimated
from X-ray \citep{1986ApJ...300..675H,2008AdSpR..41..383K} and optical 
\citep{2002A&A...391..705W} observations.  
Signatures of RS effect are suggested in a few previous X-ray spectroscopic studies of Cygnus Loop. 
With {\sl ASCA} observation, conventional NEI model revealed that the abundance of O relative to other elements in the northeast decreases at the shell region \citep{1999ApJ...525..305M}.
{\sl Chandra} observation shows that the abundance of O group elements is roughly half those of the Ne group elements (Ne, Na, Mg, Al, Si, S and Ar) and Fe group elements (Ca, Fe and Ni) at the southwestern edge \citep{2004MNRAS.351..385L}. 
{\sl Suzaku} observation of the northeastern rim \citep{2008PASJ...60..521M} inferred a low abundance of O, which is depleted by an factor of two compared with other heavy elements from C to Fe; meantime, the optical depth of \OVIIr\ line is found apparently larger than those of other resonant lines.
These results have been suggested to be indicative of the RS effect on the O\,VII\,Her line, especially along the rim of the SNR.




Recently, \citet{2019ApJ...871..234U} obtained a high O\,VII G-ratio up to $1.79\pm0.09$ for the knotty structure ``SW-K" in Cygnus Loop from {\it XMM-Newton} RGS observation.
By invoking a single NEI component with neutral absorber (the ``single NEI" model), they found a significant discrepancy between the model and the data, especially for the O\,VII\,He$\alpha$ triplet \citep[Figure\,3 in][]{2019ApJ...871..234U}, which shows that the intensities of the O\,VII\,f line and O\,VIII\,Ly$\alpha$ line are significantly underestimated.
Even an addition of a CX component (the ``NEI$+$CX" model) could not simultaneously fit the f and r lines of the O\,VII\,He$\alpha$ triplet.
By introducing 
an ionized absorber localized around the SW-K (the ``[Ionized\,Abs.$\rm\times(NEI+CX)]_{shell}$" model), 
a spectral fit was obtained (with $\nH\sim 5.5\cm^{-3}$) to explain the OVII G-ratio, which suggests the contribution of the self-absorption, or the RS, is significant for the SW-K.

\begin{figure}[ht!]
\centerline{
\includegraphics[width=.5\textwidth]{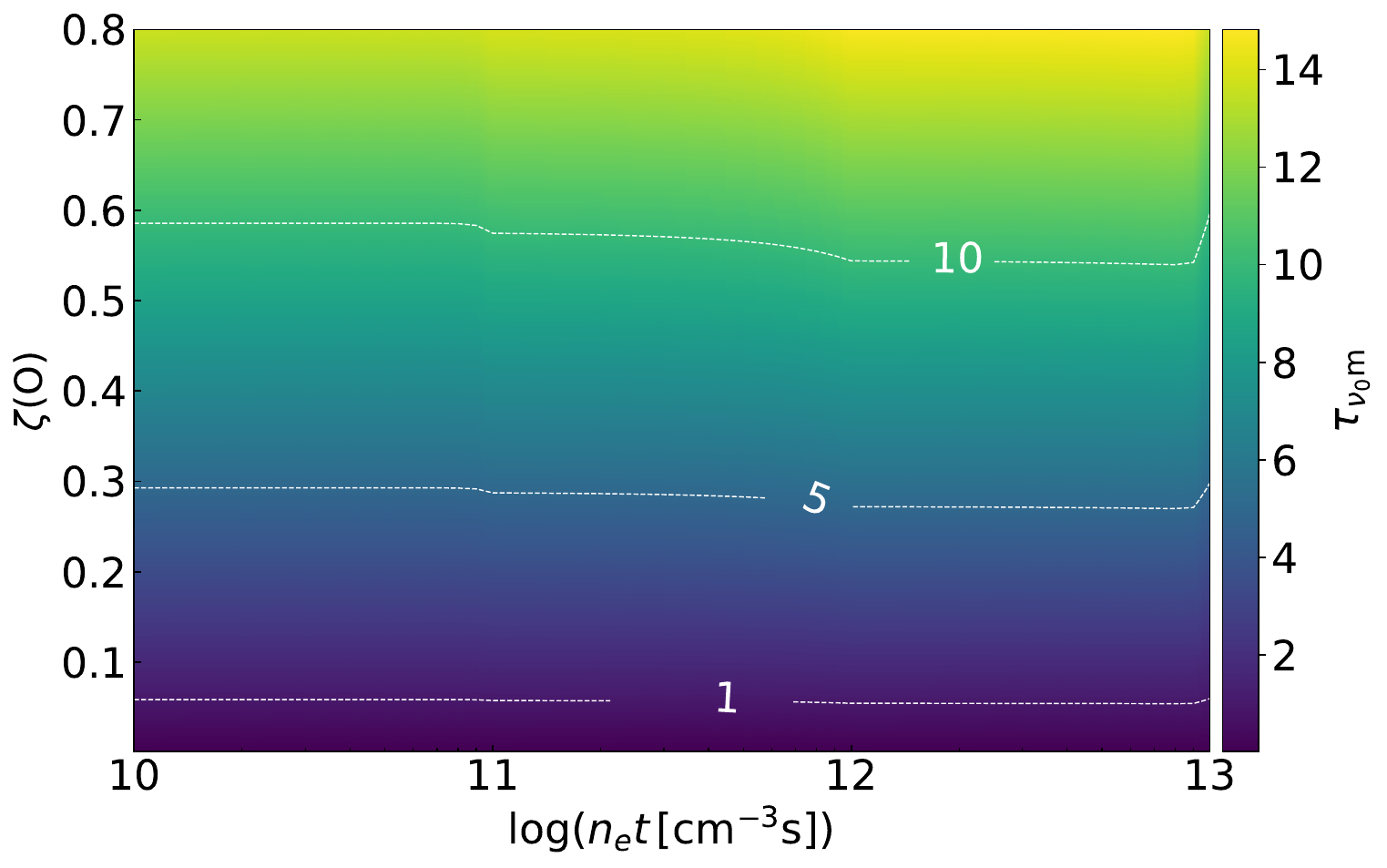}
}
\caption{Dependence of $\taum$ on $\zeta$(O) and $\nel\ti$ for Case~2.}
\label{fig:tau_net_Z}
\end{figure}

Here, we discuss the contribution of RS in the case of Cygnus Loop (especially the SW-K) based on our simulations.
First, our simulations show that the RS effect can significantly enhance the G-ratio under a similar condition of Cygnus Loop.
Figure\,\ref{fig: Gratio} shows that, even if only RS is considered, the G-ratio near the edge of the remnant could be raised to above 1 for relatively large $\taum$ in Case~2.
Furthermore, a G-ratio $\ga1.8$ as observed in SW-K can be achieved if $\taum\sim10$, which corresponds to an O abundance $\zeta({\rm O})\sim0.53$. Thereby, the observed high G-ratio of SW-K may be exclusively explained by the RS effect.
We note that, the ionization parameter makes little effect on the inference above. Specifically, in a broad range of $\nel\ti$ from $1\times10^{10}\cm^{-3}$ to $1\times10^{13}\cm^{-3}$, 
$\taum$ is greater than 1 given $\zeta$(O) above 0.1, and $\taum$ greater than 10 given $\zeta$(O) above 0.6,
as seen from Figure\,\ref{fig:tau_net_Z} which illustrates the dependence of $\taum$ on $\zeta$(O) and $\nel\ti$ for Case 2.
%
%

\begin{figure}[ht!]
\centerline{
\includegraphics[scale=0.6]{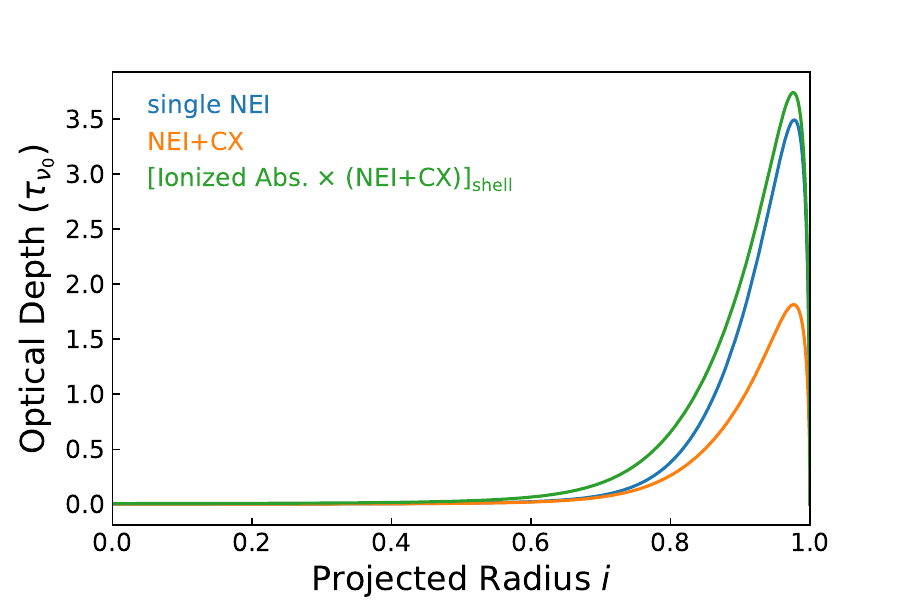}
\includegraphics[scale=0.6]{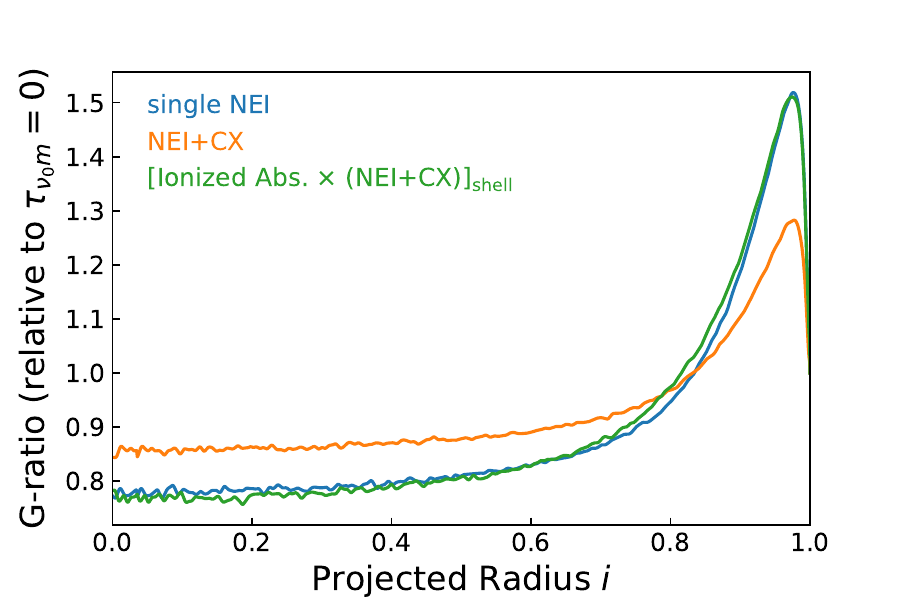}
}
\caption{Left: The distribution of the line-center optical depth of \OVIIr\ line for three models in \citep{2019ApJ...871..234U}. 
The plasma model parameters are:
$\nH=6.3\cm^{-3}$, $\nel\ti=5.2\E{10}\cm^{-3}\s$ and $\zeta({\rm O})=0.16$
for the ``single NEI" model ({\em blue}),
$\nH=7.2\cm^{-3}$, $\nel\ti=3.6\E{10}\cm^{-3}\s$ and $\zeta({\rm O})=0.069$ for the ``NEI+CX" model ({\em orange}), and
$\nH=5.5\cm^{-3}$, $\nel\ti=2.6\E{10}\cm^{-3}\s$ and $\zeta({\rm O})=0.18$
for the ``[Ionized\,Abs.$\rm\times(NEI+CX)]_{shell}$" model ({\em green}). The shock velocity $V_s=$300\,km$\ps$ and the postshock temperature $kT_s=0.2$\,keV are adopted.
Right: The distribution of G-ratio values with RS relative to those without RS.
}
\label{fig: Cy_tau}
\end{figure}

On the other hand, the contribution of the CX process probably could not be ruled out in accounting for the high G-ratio; however, even if CX is taking place, the RS effect would still be non-negligible.
It could be readily shown that, when the RS effect is included, the G-ratio values of the SW-K that were calculated from the spectral fit of ``single NEI" model and the ``NEI+CX" model have a considerable room to be increased.
The distributions of the LOS optical depth of the \OVIIr\ line across the remnant with the plasma parameters of the both models adopted are plotted in Figure\,\ref{fig: Cy_tau} (left panel). 
It illustrates that the optical depths are above 1 near the edge for the both models, which indicates that the RS effect should not be neglected there. 
The corresponding distributions of G-ratio across the remnant are obtained by simulation (Figure\,\ref{fig: Cy_tau}, right panel), which shows that the G-ratio for the two models can be increased by $\sim1.5$ and 1.3 times, respectively, near the edge. 
Namely, the G-ratios of the SW-K in the ``single NEI" model and the ``NEI+CX" model, 0.774 and 1.14 \citep[obtained from Table\,2 in][]{2019ApJ...871..234U}, would be raised to 1.2 and 1.5, respectively.
(The G-ratio distribution obtained from our simulation for the ``[Ionized\,Abs.$\rm\times(NEI+CX)]_{shell}$" model is similar to that for the ``single NEI" model.)
The observed ratio $\sim1.8$ could be achieved if the O abundance (and accordingly $\taum$) is properly increased.
Such cases, together with the ``[Ionized\,Abs.$\rm\times(NEI+CX)]_{shell}$" model which incorporates an ionized self-absorber to mimic the spectral effect of RS to some extent, demonstrate that the RS can be one of significant mechanisms in interpretting the high G-ratio even if CX plays a role in the O~VII~He$\alpha$ triplet emission. 
Thus, it is very likely that the high G-ratio in Cygnus Loop is caused by a combination of both RS and CX.

\subsection{Expectation for RS in ejecta-dominated SNRs}\label{sec:Simple discussion on RS in ejecta-dominated SNR}

Our framework of simulation for RS of X-ray lines can be extended to ejecta-dominated SNRs.
In Appendix\,\ref{app: The RS of Ejecta}, adopting some basic parameters similar to those in Cas\,A, we present simulations of RS of \OVIIr\ and OVIII\,Ly$\alpha$ lines on some simplified assumptions.
We ignore the O lines from the shocked ISM and assume that the shocked ejecta are uniform in both density and temperature.

We show that the the line-center optical depth ($\tau_{\nu_0}$) of resonant lines \OVIIr\ and \OVIIIa\ peaks at the
projected radius of the inner boundary of the shocked ejecta,
and \OVIIr\ is prominently optically thicker than \OVIIIa\ in the shocked region of the ejecta.
The \OVIIr\ line emission from the entire remnant has a broad profile (with a full width at half maximum (FWHM) $\sim12$\,eV)  
with a a saddle-like shape at the top.
Such line profiles are expected to be resolvable 
with next-generation high-resolution telescopes.
Like the case of the ST phase, some of the \OVIIr\ photons are spatially redistributed by RS from the outer regions to the inner region.
The G-ratio of the \OVIIr\ triplet is evidently enhanced in the shocked region and peaks at the projected inner boundary.
In the shocked ejecta, the RS effect on the \OVIIIa\ line is non-negligible, although with $\tau_{\nu_0}$ under 1, and the \OVIIIb/\OVIIIa\ ratio is shown to be moderately enhanced.
This is qualitatively consistent with the \OVIIIa\ deficit in the southeastern region of Cas\,A.

\section{Summary} \label{sec: Summary}
We conduct a set of MC simulations to investigate the impact of the RS effect on the soft X-ray resonant-line emission (typified by \OVIIr)  from SNRs.
The spatial distribution of the physical conditions is characterized by the Sedov-Taylor self-similar solution, and the basic parameters used for modelling are generally similar to those of Cygnus Loop.
The principal properties revealed in our simulation are summarized in the following.
\begin{enumerate}
    \item The distribution of the LOS optical depth of the scattering indicates that the impact of the RS effect is most significant near the edge of the remnant.
    \item The line profile for the outer projected region is depressed in height with the increase of the optical depth, and the line profile for the inner region takes a shape of double peaks due to Doppler effect of the expanding shell.
    \item For the inner region and the entire remnant, the double-peak line profiles can deviate from bilateral symmetry. This is a result of a sum of the scattered photons produced from the expanding hot gas at different radii with different temperatures and different photon production efficiencies.
    \item Owing to RS, the SB of the line emission is modified, with a decrease in the outer projected region and an enhancement in the inner projected region.
    \item With RS taken into account, the G-ratio of the O\,VII emission is elevated to high values in the outer region and is relatively low in the inner region. In contrast, the distribution of the O\,VIII/O\,VII flux ratio remains essentially unchanged.
    \item We employ our simulation to assess the O\,VII G-ratio of the SW-K region near the edge of Cygnus Loop.
    We find that, with the physical conditions derived in previous literature adopted, the RS effect is non-negligible. 
    This further indicates that the previous assessments of the O abundance may have been underestimated because of ignoring the RS effect.
    \item Simulation performed for the SNRs in ejecta-dominated phase exemplified by Cas\,A shows that RS in the shocked ejecta may have some apparently effects on the observational properties of oxygen resonant lines.
\end{enumerate}

In the near future, the next-generation X-ray missions such as XRISM, HUBS, LEM, and Arcus will hopefully enable us to probe the flux change of the resonant lines and even capture the modification of the line profiles.
The spatially-resolved spectroscopy study may help to map the SB of the resonant lines with comparison to other lines and may be of great importance in distinguishing RS from other mechanisms like CX.
\section{Acknowledgments}
\begin{acknowledgments}
We appreciate Ze-Cheng Zou for various helpful discussions in the study.
Y.C.\ acknowledges the support from the
NSFC under grants 12173018 and 12121003.
L.S.\ acknowledges the support from Jiangsu Funding Program for Excellent Postdoctoral Talent (2023ZB252).
G.Z.\ acknowledges the support from FONDECYT Postdoctorado through grant 3200841.
S.Z.\ acknowledges the support from the NSFC grant 12273112.
\end{acknowledgments}

%

\vspace{5mm}


\software{PyAtomDB \citep{2020Atoms...8...49F}
          }



\appendix

\section{RS in Ejecta-dominated SNRs} \label{app: The RS of Ejecta}
For the young SNRs, especially those in the free-expansion phase (or ejecta-dominated phase) as exemplified by Cas\,A,
due to high velocity of the blast wave, the temperature of the post-forward-shock gas is very high \citep[e.g., $\ga2$\,keV,][]{2014ApJ...789....7L},
so that the ionization fraction of O\,VII (as well as O\,VIII) is very low \citep[e.g.,][]{2012A&ARv..20...49V}.
In Cas\,A, actually, the thermal X-ray emission which is from shocked ISM is much fainter than from shocked ejecta \citep[e.g.][]{2006ApJ...651..250L}. 
In addition, the OVII and OVIII resonant lines there are found optically thin \citep{1995A&A...302L..13K}.
We thus, for simplicity, ignore the O resonant line emission from the shocked ISM in our simulations of RS of \OVIIr\ and \OVIIIa\ lines of the SNRs in the ejected-dominated phase.

We assume that the shocked ISM is uniformly distributed within a shell, with a depth of $\Rs/12$, outside the contact discontinuity,
and the shocked ejecta are uniformly distributed between the reverse shock and the contact discontinuity. 
Hereafter, our simulation for the RS in the ejecta-dominated phase will be performed with parameters similar to those in Cas\,A.
According to \citet{1999ApJS..120..299T}, we have the outer radius of the remnant (i.e., the forward shock) 
$\Rs\approx2.2$\,pc  (at a distance 3.4\,kpc) and the radius of the reverse shock 
$R_{\rm r}\approx1.6$\,pc  for the case with the ejecta distribution index $n=2$ and a uniform ambient medium;
here we have accordingly adopted other parameters:
total ejecta mass $M_{\rm ej}=3M_{\odot}$, ambient medium density $n_0=3.2 \cm^{-3}$, explosion energy $E=10^{51} \erg$, 
a reverse shock velocity in the frame of the unshocked ejecta $\tilde{V}_{\rm r}\sim3\times10^3\km\s^{-1}$,
and a modified age 350\,yr.
The oxygen mass in the shocked ejecta
is thus
$\sim 0.36 M_{\odot}$ if we adopt the oxygen mass fraction in the ejecta $\sim60\%$ \citep{1996A&A...307L..41V}. 
Hence, the oxygen ion density therein is
$n_{\rm O}\sim0.05\cm^{-3}$.

The O\,VII and O\,VIII ion density can be obtained from
$n_{\rm ion}=n_{\rm O}f_{\rm ion}(T_{\rm e})$,
where $T_{\rm e}$ is the electron temperature in the shocked ejecta.
Considering electron-proton temperature non-equilibration as usually found in young SNRs,
we adopt an electron temperature $T_{\rm e}=1$\,keV for the shocked ejecta with reference of the post-forward-shock electron temperature of SN\,1006 where the shock velocity ($\sim2.8\E{3}\km\ps$) is similar to $\tilde{V}_{\rm r}$ adopted here \citep{2023ApJ...949...50R}.
We also assume that the oxygen ions have not sufficiently exchanged thermal energies with other ion species,
and hence have
the temperature of the oxygen ions
$T_{\rm ion}=(3/16) \mu_a m_{\rm p}\tilde{V}^2_{r}/k
  \sim 280\keV$, 
with the atomic weight $\mu_a=16$ for oxygen ions,
and a thermal velocity dispersion of the ions $\sigma_v\sim 1.3\E{3} \km\s^{-1}$.
At this temperature, the oxygen ions comprise mainly O\,VII and O\,VIII
with fractions 
$f_{\rm ion}\approx 0.46$ and 0.47, respectively. 

With the aforementioned elemental and ionic mass fractions, the mean election number density in the shocked ejecta is estimated as 
$\nel\sim0.9\cm^{-3}$.
With the remnant age used for approximation, we estimate the ionization parameter as $\nel\ti\sim1\times10^{10} \cm^{-3}\s$.

We further assume that the radial velocity of the bulk expanding motion is linearly distributed between the reverse shock 
and the contact discontinuity. 
With the velocities of the reverse shock in the ambient rest frame $1580\km\ps$ \citep{1999ApJS..120..299T} and in the frame of the unshocked ejecta $\tilde{V}_{\rm r}$,
we have the expansion velocity of the gas at the inner boundary of the shocked ejecta $\sim2330\km\ps$.
The expansion velocity at the outer boundary of the shocked ejecta is approximated as $(11/12)\Vs\sim3300\km\ps$.
 
\begin{figure}[ht!]
\centerline{
\subfigure[]{
    \includegraphics[scale=0.6]{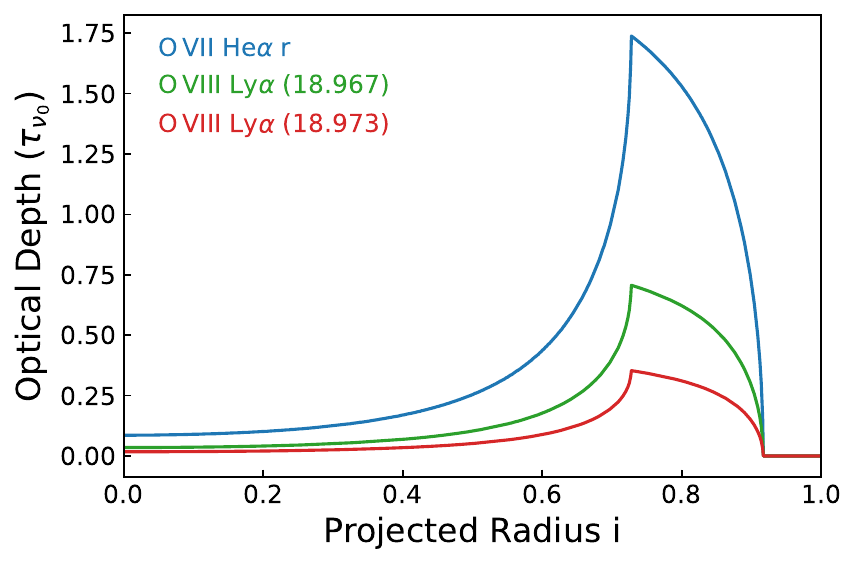}
    \label{fig:tau_CasA}
}
\subfigure[]{
    \includegraphics[scale=0.6]{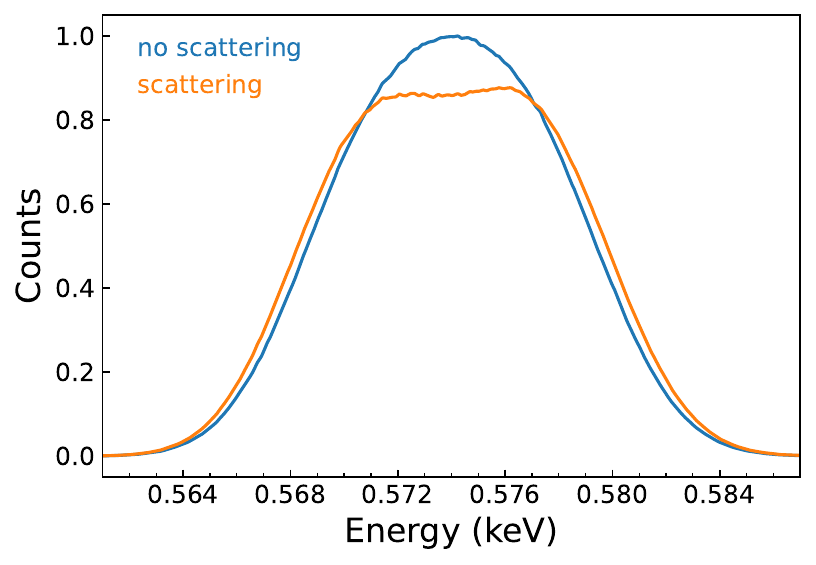}
    \label{fig:OVII_CasA}
}
}
\caption{(a): The distribution of the line-center optical depth of \OVIIr\ and O\,VIII\,Ly$\alpha$ modeled for the young SNR. (b): 
Line profiles of \OVIIr\ photons with and without RS modeled for the entire young remnant.}
\end{figure}

Using Eq.(\ref{eq: tau}), we calculate the projected radial distribution of the line-center optical depth ($\tau_{\nu_0}$) of \OVIIr\ and OVIII\,Ly$\alpha$ lines in the SNR, as plotted in Figure\,\ref{fig:tau_CasA}.
As seen in the figure, $\tau_{\nu_0}$ of the O resonant lines peaks at the projected radius of the inner boundary of the shocked ejecta.
The \OVIIr\ line is optically thick (with $\tau_{\nu_0}$ higher than 1) in the range 
$0.7\la i \la 0.9$, overlapping with most part of 
the shocked ejecta.
The optical depth $\tau_{\nu_0}$ of the O\,VIII\,Ly$\alpha$ (18.967\AA) is close to 
0.75 in the projected shocked region, and therefore the RS it suffers may sometimes be non-negligible there,
while $\tau_{\nu_0}$ of the O\,VIII\,Ly$\alpha$ (18.973\AA) is relatively low across the remnant. 

The line profile of \OVIIr\ emission subjected to resonant scattering in the SNR is calculated and plotted in Figure\,\ref{fig:OVII_CasA}.
In the case without RS effect, the line profile is broad (with a FWHM $\sim11$\,eV) due to large thermal velocity dispersion and Doppler shifts, but does not appear flat at the top as those expected for the ST phase (see \S\ref{sec:lineprofile}). 
With RS, the line profile is diminished around the line center and moderately broadened (by 
$\sim0.5$\,eV in each side) with a saddle-like shape.
In this circumstance, even after convolving the instrumental response, the scattering-induced line broadening might be detectable 
with next-generation high-resolution telescopes such as HUBS and LEM.

\begin{figure}[ht!]
\centerline{
\subfigure[]{
    \includegraphics[scale=0.6]{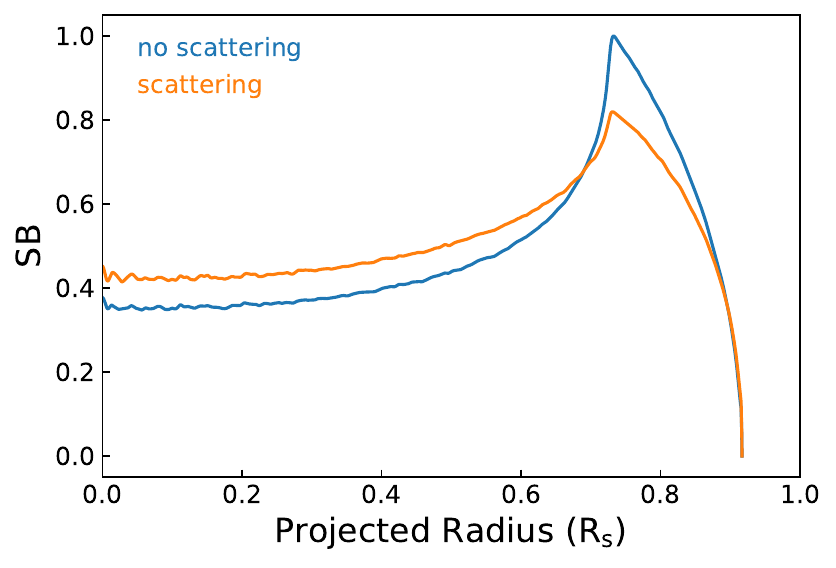}
    \label{fig:OVIISB_CasA}
    }
\subfigure[]{
    \includegraphics[scale=0.6]{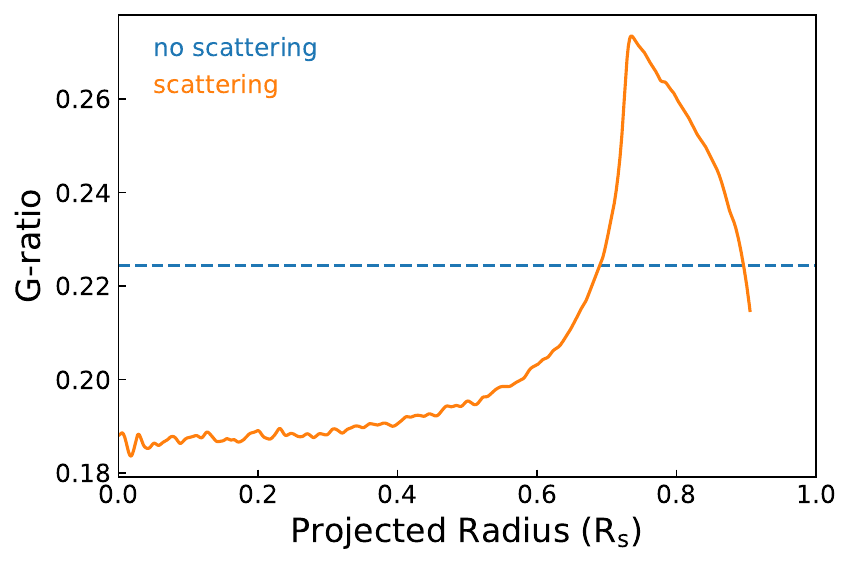}
    \label{fig:Gratio_CasA}
}
}
\caption{(a): Projected radial SB distribution of the O\,VII He$\alpha$ r line with and without RS for the young SNR. (b): Projected radial profiles of the O\,VII He$\alpha$ triplet G-ratio 
modeled for the young SNR.}
\end{figure}

The projected radial SB distribution of the \OVIIr\ line is shown in Figure\,\ref{fig:OVIISB_CasA}.
It shows that line emission brightness in the shocked ejecta is decreased by RS, while the brightness in the inner region is elevated so that the contrast in surface brightness between the projected inner and outer regions is lessened.

Figure\,\ref{fig:Gratio_CasA} shows that the G-ratio of the \OVIIr\ triplet is expected to be evidently enhanced in the projected region of the shocked ejecta and peak at the projected inner boundary. 

\begin{figure}[ht!]
\centering
    \includegraphics[scale=0.6]{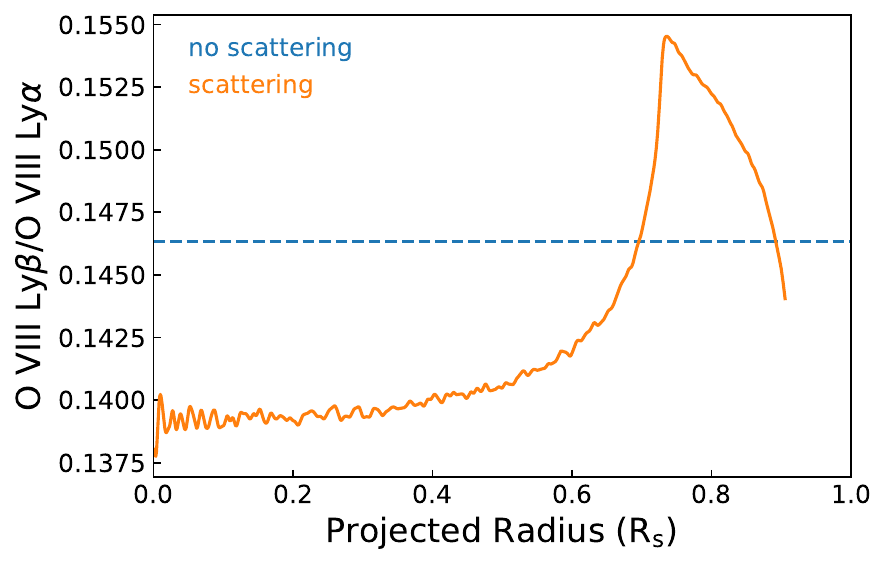}
\caption{Projected radial profile of the O\,VIII\,Ly$\beta$/O\,VIII\,Ly$\alpha$ ratio modeled for the young SNR.}\label{fig:Lyratio_CasA}
\end{figure}

We also calculate projected radial distribution of the line ratio between O\,VIII\,Ly$\beta$ and O\,VIII\,Ly$\alpha$ (Ly$\beta$ has a rather small oscillation strength, not subjected to RS).
As shown in Figure\,\ref{fig:Lyratio_CasA}, the ratio has a similar variation trend as that of the G-ratio, enhanced in the shocked ejecta due to the suppression of Ly$\alpha$ flux, but the enhancement is moderate.
\citet{2001A&A...365L.225B} reported that the O\,VIII\,Ly$\alpha$/Ly$\beta$ ratio decreases from $\sim0.5$ to $\sim0.15$--0.2 along the NW-SE direction of Cas\,A, indicating a deficit of Ly$\alpha$ in the SE region of the remnant.
They suggested some possible reasons affecting the line ratio including the line blending (i.e., O\,VIII\,Ly$\beta$ and Fe-L lines), variation of absorbing hydrogen column $N_{\rm H}$, and RS.
It was suggested that RS be one of the possible origins of the Ly$\alpha$ deficit. 
Our results thus are in favor of the role that RS effect may play in the Ly$\beta$/Ly$\alpha$ enhancement due to Ly$\alpha$ deficit in the SE region, if this region is dominated by shocked O-rich ejecta.
Also, if the oxygen ion density in the southeastern ``blob"  \citep{2001A&A...365L.225B} is higher due to inhomogenity than that used in our calculation, higher optical depth of \OVIIIa\ and stronger RS of it can be expected.

\bibliography{rs}{}
\bibliographystyle{aasjournal}

\end{document}